\documentclass[aps,twocolumn,nofootinbib]{revtex4}
\usepackage{graphicx}
\usepackage{amsmath}
\usepackage{amssymb}
\usepackage{mathtools}
\usepackage{hyperref}
\usepackage{xcolor}
\usepackage{bm}
\usepackage[T1]{fontenc}
\usepackage{upquote}

\newcommand{\rn}{{\rm n}}

\begin{document} 

\title{\bf Eikonal solutions for moment hierarchies of Chemical
  Reaction Networks in the limits of large particle number}

\author{Eric Smith}

\author{Supriya Krishnamurthy}

\date{\today}

\begin{abstract}

Trajectory-based methods are well-developed to approximate
steady-state probability distributions for stochastic processes in
large-system limits.  The trajectories are solutions to equations of
motion of Hamiltonian dynamical systems, and are known as
eikonals. They also express the leading flow lines along which
probability currents balance.  The existing eikonal methods for
discrete-state processes including chemical reaction networks are
based on the Liouville operator that evolves generating functions of
the underlying probability distribution.  We have previously
derived~\cite{Krishnamurthy:CRN_moments:17,Smith:CRN_moments:17} a
representation for the generators of such processes that acts directly
in the hierarchy of moments of the distribution, rather than on the
distribution itself or on its generating function.  We show here how
in the large-system limit the steady-state condition for that
generator reduces to a mapping from eikonals to the ratios of
neighboring factorial moments, as a function of the order $k$ of these
moments.  The construction shows that the boundary values for the
moment hierarchy, and thus its whole solution, are anchored in the
interior fixed points of the Hamiltonian system, a result familiar
from Freidlin-Wenztell theory.  The direct derivation of eikonals from
the moment representation further illustrates the relation between
coherent-state and number fields in Doi-Peliti theory, clarifying the
role of canonical transformations in that theory.

\end{abstract}

\maketitle

\section{Introduction: directly solving for fluctuation moments in
stochastic Chemical Reaction Networks}

When solving for the steady states of stochastic processes,
approximation methods are needed in all but the simplest cases.  The
choice of representation for the process and its generator can dictate
the efficiency or comprehensibility of approximation methods that are
available in that representation.

Two representations for the generators of stochastic processes that
are now more than a century old~\cite{vanKampen:Stoch_Proc:07} are the
transition matrix of the master equation, acting directly on
probability distributions, and the generator of the Liouville
equation, acting on generating functions that are the Laplace
transforms of those distributions.

A third representation can be constructed for the generator acting
directly on the hierarchy of moments. For a class of discrete-state
stochastic processes, we have obtained an explicit form for such a
generator acting on the moment
hierarchy~\cite{Krishnamurthy:CRN_moments:17,Smith:CRN_moments:17}.
These are the models of stochastic Chemical Reaction Networks (CRNs),
which have been extensively
studied~\cite{Feinberg:def_01:87,Anderson:product_dist:10,Gunawardena:CRN_for_bio:03,Baez:QTRN_eq:14,Polettini:open_CNs_I:14,Polettini:diss_def:15}
and are of both mathematical and practical interest.  The moments of a
probability distribution for a CRN arrange into a lattice with
coordinates $k$, where $k$ is a vector whose components determine the
moment of interest.

For CRNs with only finitely many reactions, the generator in the
moment lattice takes the form of a discrete Laplacian, and is solved
analogously to a heat equation.  Because the dynamical equation for
moments is linear, the steady-state condition can be reduced to a
collection of constraints, one per lattice site.  Time-invariance of
the moment at index $k$ implies a constraint on the \emph{ratios} of
neighboring moments to the moment at $k$.

If a CRN has only one dynamical degree of freedom, a discrete
Laplacian acting on its moments reduces to a recursion relation on the
moment ratios.  We showed
in~\cite{Krishnamurthy:CRN_moments:17,Smith:CRN_moments:17} that such
recursions can be solved by a method of matched asymptotic expansions,
in which low-order modes are solved recursively as one would expect,
upward from the mean or variance, but high-order modes are solved
(counterintuitively) by anchoring the moment ratios at large $k$,
where they converge asymptotically to values set by parameters in the
generator.  For multivariate distributions, however, the corresponding
method of solution encounters the difficulty faced in the solution of
other Laplacian equations, which is the specification of boundary
conditions.  For the CRN moment hierarchies, the difficult boundaries
are those on which one or more of the moments is \emph{not} of
asymptotically large order, since there is no approximation scheme
which gives an estimate of such moments.

An alternative to direct solution of Laplacian equations, which
addresses the problem of boundary conditions both computationally and
conceptually, is the WKB method, which computes the leading
exponential dependence of probability distributions in large-system
limits.  The key quantities in the WKB approximation are eikonals:
locally least-improbable trajectories for improbable excursions, along
which balance of probability currents determines the large-deviation
behavior of the
distribution.\footnote{See~\cite{Touchette:large_dev:09} for the
explanation that large-deviations scaling is the separation of scale
from structure in probability distributions for aggregate statistics.
Trajectories of eikonals are properties of the large-deviation rate
functions, while the probabilities associated with them are
exponential in the scale factors.}  Eikonal solutions for stationary
distributions, boundary values, and first-passage statistics of
diffusion solutions in basins of attraction were extensively developed
by Freidlin and Wentzell~\cite{Freidlin:RPDS:98}.  Related
formulations in terms of 2-field functional integrals, for stochastic
processes on discrete state spaces including those of CRNs, were
derived by Doi and
Peliti~\cite{Doi:SecQuant:76,Doi:RDQFT:76,Peliti:PIBD:85,Peliti:AAZero:86}.

The stability properties of WKB methods are dual to those of the
moment recursions
in~\cite{Krishnamurthy:CRN_moments:17,Smith:CRN_moments:17}, in that
they propagate probability
distributions outward into attenuating tails from the regions near the
mean where probability is most concentrated.  Those central regions
are the main scale factors in the recursion approach, where matching
conditions must be imposed between low and high moment orders.
It is by grounding asymptotic regions in the centers of mass of probability
distributions that WKB methods provide approximate solution methods
for boundary-value problems~\cite{Freidlin:RPDS:98}.

In this paper we derive the eikonal method to directly approximate the
ratios of adjacent moments in the moment hierarchy.  Our starting
point will be the generator
from~\cite{Krishnamurthy:CRN_moments:17,Smith:CRN_moments:17}.  The
steady-state condition will reduce to a set of stationary-path
equations that are the same as those for eikonals in Doi-Peliti
theory.  The new feature resulting from the moment representation is a
mapping from the order $k$ of a moment to a specific position along
the particular escape trajectory that governs the ratio of that moment
to its local neighbors.

The analysis leads to three results: 1) The generator in the moment
representation reduces in the large-system limit to the Liouville
operator from the generating-function representation.  2) Boundary
solutions for the WKB approximation are anchored on a discrete set of
interior fixed points in the steady state, as they are in
Freidlin-Wentzell theory.  3) The reduction of the generator from the
moment representation to the Liouville operator clarifies the roles of
number fields relative to coherent-state fields in Doi-Peliti theory,
as these represent both activities and relative migration rates in
steady states. 

The presentation is organized as follows: Sec.~\ref{sec:three_reps}
reviews three representations of the generator for a population
process such as a CRN: the transition matrix of the master equation,
the Liouville operator, and the generator in the moment
representation.  Sec.~\ref{sec:eikonal_solns} derives the eikonal
approximation for the moment hierarchies of CRNs.  The large-number
limit of the steady-state recursions of the moment hierarchy is
derived, recovering the Liouville operator. It is then shown how the
orders of the moments map to positions on the eikonals of the
Liouville represenation.  Sec.~\ref{sec:examples} illustrates the
eikonal construction on three example CRNs, beginning with a model
that is exactly solvable to demonstrate convergence and show how
finite-size effects are handled, and culminating in a model where
eikonal methods are needed, but where they also break down in some
regions due to trajectory crossing.

\section{Representations of the generator for discrete-state
stochastic processes}
\label{sec:three_reps}

To begin, we specify and introduce notation for the class of
stochastic processes to be considered, and review the three
representations of the generator.  We will consider continuous-time
Markov processes on discrete state spaces that are lattices of
non-negative integers in some dimension $P$.  Examples include a
variety of population processes~\cite{Smith:evo_games:15}, among which
are stochastic CRNs described in more detail below.

States are indexed by vectors $\rn \equiv \left[ {\rn}_p \right]$, in
which $p \in 1 , \ldots , P$ are called \textit{species} and ${\rn}_p$
is a non-negative integer giving the count (particles, individuals,
etc.) of species $p$ in the population.  Probability distributions
over states $\rn$ are denoted ${\rho}_{\rn}$.  A time-dependent
probability evolves under a master equation of the form
\begin{equation}
  \frac{d}{d\tau}
  {\rho}_{\rn} = 
  \sum_{{\rn}^{\prime}}
  {\rm T}_{\rn {\rn}^{\prime}}
  {\rho}_{{\rn}^{\prime}} . 
\label{eq:master_eq_popproc}
\end{equation}
${\rm T}_{\rn {\rn}^{\prime}}$ is a left-stochastic
matrix\footnote{Left-stochastic means that $\sum_{\rn} {\rm T}_{\rn
{\rn}^{\prime}} = 0, \; \forall {\rn}^{\prime}$, so that
Eq.~(\ref{eq:master_eq_popproc}) conserves probability.} called the
\textit{transition matrix}, and is one representation of the
generator.

\subsection{Moment-generating function and Liouville operator}

The Laplace transform of a distribution ${\rho}_{\rn}$ is called its
moment-generating function (MGF).  While the analytic structure of the
Laplace transform is often a source of elegant solution methods and
the reason to use the generating
function~\cite{Flajolet:anal_comb:09}, many problems require only its
formal representation as a power series, and the ensuing linear
algebra.  For these, an abstract operator algebra such as that of
Doi~\cite{Doi:SecQuant:76,Doi:RDQFT:76} is more economical, especially
for evaluating quadratures of Eq.~(\ref{eq:master_eq_popproc}).  The
Doi algebra is reviewed in numerous
places~\cite{Mattis:RDQFT:98,Kamenev:DP:02,Smith:LDP_SEA:11,Baez:QTRN:13,Baez:QTSM:17},
and we provide only a summary here.

\textit{Lowering operators} $a_p$ that remove particles from a
population state, and \textit{raising operators} $a^{\dagger}_q$ that
introduce particles, are defined by their commutation relation
\begin{equation}
  \left[
    a_p , a^{\dagger}_q
  \right] = 
  {\delta}_{pq} , 
\label{eq:ladd_comm}
\end{equation}
where $p , q \in 1 , \ldots , P$ and ${\delta}_{pq}$ is the Kronecker
$\delta$ symbol.

\textit{Number states}, corresponding to point distributions at each
lattice index $\rn$, are defined by the action of the lowering and
raising operators on them, as 
\begin{align}
  a_p \left| \rn \right)
& = 
  {\rn}_p \left| \rn - 1_p \right)
& 
  a^{\dagger}_p \left| \rn \right)
& = 
  \left| \rn + 1_p \right) . 
\label{eq:numb_ladd_def}
\end{align}
It follows that all number states can be built up from a ground state
$\left| 0 \right)$ by multiplication with suitable powers of the
$\left\{ a^{\dagger}_p \right\}$.  

The MGF of a distribution ${\rho}_{\rn}$ is the sum of number states
weighted by the coefficients of ${\rho}_{\rn}$, denoted 
\begin{equation}
  \left| \phi \right) = 
  \sum_{\rn}
  {\rho}_{\rn}
  \left| \rn \right) . 
\label{eq:Doi_state_n_basis}
\end{equation}
Eq.~(\ref{eq:master_eq_popproc}) implies a corresponding evolution
equation for the MGF, known as the Liouville equation, 
\begin{equation}
  \frac{d}{d\tau}
  \left| \phi \right) = 
  - \mathcal{L} \! \left( a^{\dagger} , a \right)
  \left| \phi \right) 
\label{eq:Liouville_eq}
\end{equation}
in which the \textit{Liouville operator} $\mathcal{L} \! \left(
a^{\dagger} , a \right)$ is derived from the matrix of values ${\rm
T}_{\rn {\rn}^{\prime}}$. 

Computing the trace of a probability distribution is represented in
the operator algebra in terms of an inner product with a dual state
$\left( 0 \right|$ (which has the interpretation of a projection
operator).  Under an inner product called the Glauber norm, 
\begin{equation}
  \left( 0 \right| 
  e^{\sum_p a_p}
  \left| \rn \right) = 
  1 \; ; \; \forall \rn ,
\label{eq:Glauber_norm}
\end{equation}
in which all number states are normalized, the generating functions
become vectors in a Hilbert space, completing the definition of the
linear algebra.  For some applications, including the Peliti
functional integral introduced below, it remains useful to express the
generating function as a function of a vector $z \equiv \left[ z_p
\right]$ of complex numbers.  The analytic representation is extracted
from the formal power series by the inner product
\begin{equation}
  \left( 0 \right|
  e^{\sum_p z_p a_p}
  \left| \phi \right) = 
  \phi \! \left( z \right) . 
\label{eq:Doi_to_anal}
\end{equation}

\subsubsection{Stochastic CRNs; the form of their Liouville operator} 

The subset of population processes that we will consider in detail are
the stochastic CRNs.  Each elementary event is an instance of some
\textit{reaction}, which removes one set of particles from the
population and places another set of particles into it.  The number of
particles of each type that are removed or added give the
stoichiometry of the reaction, and the rate of the reaction equals the
probability for an instance to occur per unit time.  Particles are
sampled proportionally and without replacement, leading to mass-action
kinetics at large numbers.

In a decomposition of CRNs due to Horn and
Jackson~\cite{Horn:mass_action:72} and
Feinberg~\cite{Feinberg:def_01:87}, the sets of particles either
removed or {\textcolor{red} created} are called \textit{complexes}.
We denote the complexes that appear in a CRN with subscripts $i$, $j$;
each reaction corresponds to a unique ordered pair $\left( ji \right)$
that removes complex $i$ and adds complex $j$.

The reaction network is specified by an adjacency matrix
${\mathbb{A}}_{\kappa}$ on the set of complexes, with the form
\begin{equation}
  {\mathbb{A}}_{\kappa} \equiv 
  \sum_{\left( i , j \right)}
  \left( w_j - w_i \right)
  {\kappa}_{ji}
  w_i^T . 
\label{eq:CRN_compl_adj}
\end{equation}
$w_i$, a column vector, is an indicator function with 1 in the $i$th
position and zero elsewhere; $w_i^T$ is its transpose.
${\kappa}_{ji}$ is the {\textcolor{red} reaction} rate constant for
reaction $\left( ji \right)$, with respect to units $d\tau$ for time.
${\mathbb{A}}_{\kappa}$ is left-stochastic on complexes, as ${\rm
T}_{\rn {\rn}^{\prime}}$ is on states.

The matrix ${\mathbb{A}}_{\kappa}$ acts to the right on a column
vector of activities for complexes.  These activities are extracted
from a state $\left| \phi \right)$ by a vector ${\psi}_Y \equiv \left[
{\psi}_Y^i \right]$ in which the components are the operators 
\begin{equation}
  {\psi}_Y^i \! \left( a \right) \equiv 
  \prod_p 
  a_p^{y^i_p}
\label{eq:Psi_Yi_def}
\end{equation}
$Y \equiv \left[ y^i_p \right]$ is the matrix of stoichiometric
components, giving the counts of species $p$ in each complex $i$.  

The {\textcolor{blue} {rate}} equation for the CRN defined by
${\mathbb{A}}_{\kappa}$ and $Y$ can be written in the form 
\begin{equation}
  \frac{dn}{d\tau} = 
  Y 
  {\mathbb{A}}_{\kappa}
  {\psi}_Y \! \left( n \right)   
\label{eq:CRN_firstmom_EOM}
\end{equation}
Here $n$ is a vector that corresponds 
in the mean-field approximation
to the vector of expectations $\left< \rn \right> \equiv \sum_{\rn}
\rn {\rho}_{\rn}$.  

As we show
in~\cite{Krishnamurthy:CRN_moments:17,Smith:CRN_moments:17}, the
Liouville operator for this class of CRNs takes the compact form 
\begin{equation}
  - \mathcal{L} \! \left( a^{\dagger} , a \right) = 
  {\psi}^T_Y \! \left( a^{\dagger} \right)
  {\mathbb{A}}_{\kappa} \, 
  {\psi}_Y \! \left( a \right)
\label{eq:L_of_as}
\end{equation}
The exact equation for the time evolution of all moments in the
distribution ${\rho}_{\rn}$, follows from this form and from the
commutation relations~(\ref{eq:ladd_comm}).

\subsubsection{The Doi-Peliti functional integral and its action
functional}

Peliti~\cite{Peliti:PIBD:85,Peliti:AAZero:86} introduced a
path-integral representation for the quadrature of the Liouville
equation~(\ref{eq:Liouville_eq}) based on an expansion in
\textit{coherent states}.  For $\xi \equiv \left[ {\xi}_p \right]$ a
vector of complex coefficients, the coherent state 
\begin{equation}
  \left| \xi \right) \equiv 
  e^{
    \left( a^{\dagger} - 1 \right) \xi
  } 
  \left| 0 \right) 
\label{eq:CS_right_def}
\end{equation}
is an eigenstate of the lowering operators, with eigenvalues
\begin{equation}
  a_p \left| \xi \right) = 
  {\xi}_p \left| \xi \right) . 
\label{eq:CS_ladd_eigstate}
\end{equation}
In Eq.~(\ref{eq:CS_right_def}) we introduce a convention that
$a^{\dagger}$ is a row vector, $\xi$ is a column vector, and
$a^{\dagger} \xi$ is the vector inner product of the two.

Through constructions that are now
standard~\cite{Smith:LDP_SEA:11,Krishnamurthy:CRN_moments:17,Smith:CRN_moments:17},
the time-dependent solution to Eq.~(\ref{eq:Liouville_eq}), converted
to complex arguments by Eq.~(\ref{eq:Doi_to_anal}), may be written as
the 2-field functional integral
\begin{equation}
  {\phi}_T \! \left( z \right) = 
  \int_0^T 
  \mathcal{D} {\xi}^{\dagger}
  \mathcal{D} {\xi}
  e^{
    \left( z - {{\xi}^{\dagger}}_T \right) {\xi}_T 
  }
  e^{- S}
  {\phi}_0 \! 
  \left( 
    {\xi}_0^{\dagger}
  \right) . 
\label{eq:DP_2FFI_CS}
\end{equation}
${\xi}^{\dagger}$ and $\xi$ are now histories indexed by time $\tau \in
\left[ 0 , T \right]$.  ${\phi}_0$ is the generating function for the
initial distribution at $\tau = 0$, and ${\phi}_T$ is its image evolved
to $\tau = T$.

All influence from the CRN dynamics in the
integral~(\ref{eq:DP_2FFI_CS}) comes through the action functional
\begin{equation}
  S = 
  \int_0^T d\tau 
  \left\{ 
    - \left( d_{\tau} {\xi}^{\dagger} \right)
    \xi + 
    \mathcal{L} \! \left( {\xi}^{\dagger} , \xi \right)
  \right\} .
\label{eq:S_from_L_CS}
\end{equation}
$\xi$ appears as a coordinate variable, ${\xi}^{\dagger}$ is its
conjugate momentum, and the place of the Hamiltonian is filled by the
Liouville function $\mathcal{L}$ in which $\xi$ takes the place of
operator $a$ and ${\xi}^{\dagger}$ takes the place of $a^{\dagger}$.
For the generator class~(\ref{eq:L_of_as}) of CRNs, we have 
\begin{equation}
  - \mathcal{L} \! \left( {\xi}^{\dagger} , \xi \right) = 
  {\psi}^T_Y \! \left( {\xi}^{\dagger} \right)
  {\mathbb{A}}_{\kappa} \, 
  {\psi}_Y \! \left( \xi \right) . 
\label{eq:L_of_xis}
\end{equation}

\subsubsection{Canonical transformations: coherent-state and
number-potential coordinates}

In terms of the coherent states used to construct the functional
integral~(\ref{eq:DP_2FFI_CS}), the components of the number index
$\rn$ correspond to a bilinear form $n_p = {\xi}^{\dagger}_p {\xi}_p$.
The number field may be made an elementary degree of freedom as part
of a canonical transformation~\cite{Goldstein:ClassMech:01} of the
coordinates of the Hamiltonian dynamical system, 
\begin{align}
  {\xi}_p^{\dagger}
& \equiv 
  e^{{\eta}_p}
& 
  {\xi}_p
& \equiv 
  e^{-{\eta}_p}
  n_p .
\label{eq:xis_to_n_eta}
\end{align}
$\eta$, the conjugate momentum to the number coordinate, admits a
variety of interpretations.  In the construction of the equilibrium
free energy of a chemical mixture, it has the dimensions of a chemical
potential~\cite{Smith:LDP_SEA:11}, so we will refer to $\left( \eta ,
n \right)$ as number-potential coordinates.  Both $\xi$ and $n$ have
dimensions of species numbers, and one of the subtleties of Doi-Peliti
theory is understanding the difference of their meaning in WKB-type
approximations.  The eikonal approximation for moment ratios is
interesting because it establishes a relation between the two in the
same observable.

The Doi-Peliti integral in number-potential coordinates becomes
\begin{equation}
  {\phi}_T \! \left( z \right) = 
  \int_0^T 
  \mathcal{D} \eta
  \mathcal{D} n
  e^{
    \left( z e^{-{\eta}_T} - 1 \right) n_T 
  }
  e^{- S}
  {\phi}_0 \! 
  \left( 
    e^{{\eta}_0}
  \right) , 
\label{eq:DP_2FFI_neta}
\end{equation}
in which the action is 
\begin{equation}
  S = 
  \int_0^T d\tau 
  \left\{ 
    - \left( d_{\tau} \eta \right)
    n + 
    \mathcal{L} \! \left( \eta , n \right)
  \right\} . 
\label{eq:S_from_L_neta}
\end{equation}
Preservation of the form of the kinetic term as $- \left( d_{\tau}
{\xi}^{\dagger} \right) \xi = - \left( d_{\tau} \eta \right) n$
ensures that Eq.~(\ref{eq:DP_2FFI_neta}) is canonical.  The Liouville
function $\mathcal{L} \! \left( \eta , n \right)$ is the same as
$\mathcal{L} \! \left( {\xi}^{\dagger} , \xi \right)$ in
Eq.~(\ref{eq:S_from_L_CS}) evaluated in the transformed variables.

\subsection{Moment hierarchy and its generator}

For CRNs with mass action rates, the macroscopic expression of
proportional sampling without replacement at the microscale, the
moments that have a simple behavior are neither ordinary moments nor
cumulants, but rather the expectations of falling
factorials~\cite{Baez:QTRN:13}, which we term \textit{factorial
moments}.  A compact notation for the falling factorial of number
$n_p$ at order $k_p$ is
\begin{align}
  {\rn}_p^{\underline{k_p}} 
& \equiv 
  \frac{
    {\rn}_p !
  }{
    \left( {\rn}_p - k_p \right) ! 
  }
& ; \; k_p \le {\rn}_p 
\nonumber \\ 
& \equiv 
 0 
& ; \; k_p > {\rn}_p .
\label{eq:factorial_moment_not}
\end{align}

In a CRN, the activity products corresponding to the complexes $i$ in
a state $\rn$ make up a column vector that we denote ${\Psi}_Y \equiv
\left[ {\Psi}_Y^i \right]$ with
\begin{equation}
  {\Psi}_Y^i \! \left( \rn \right) \equiv 
  \prod_p 
  {\rn}_p^{\underline{y_p^i}} 
\label{eq:Psi_Y_def}
\end{equation}
In the Liouville equation~(\ref{eq:Liouville_eq}), the
complex-lowering operators ${\psi}_Y^i$ from Eq.~(\ref{eq:Psi_Yi_def})
act respectively on number states and coherent states as
\begin{align}
  {\psi}_Y^i \! \left( a \right) 
  \left| \rn \right) 
& = 
  {\Psi}_Y^i \! \left( \rn \right)
  \left| \rn - y^i \right) 
\nonumber \\ 
  {\psi}_Y^i \! \left( a \right) 
  \left| \xi \right) 
& = 
  {\psi}_Y^i \! \left( \xi \right) 
  \left| \xi \right) . 
\label{eq:Psi_psi_on_n_xi}
\end{align}
Finally, in parallel to the designation $\phi \! \left( z \right)$
from Eq.~(\ref{eq:Doi_to_anal}) for the MGF, we denote the lattice of
factorial moments $\Phi \equiv \left[ {\Phi}_k \right]$ with
\begin{equation}
  {\Phi}_k \equiv 
  \left<
    \prod_p
    {\rn}_p^{\underline{k_p}}
  \right> . 
\label{eq:Phi_def}
\end{equation}

We showed in~\cite{Krishnamurthy:CRN_moments:17,Smith:CRN_moments:17}
that the Liouville equation~(\ref{eq:Liouville_eq}), used to evolve
the probability distribution defining the average in
Eq.~(\ref{eq:Phi_def}), implies the equation of motion for the moment
hierarchy
\begin{align}
\lefteqn{
  \frac{d}{d\tau} {\Phi}_k = 
  \dot{\prod_p}
  \left[ 
    \sum_{j_p = 0}^{k_p}
    \left(
      \begin{array}{c}
        k_p \\ j_p 
      \end{array}
    \right)
    Y_p^{\underline{j_p}} \,  
    e^{
      {\left( k - j \right)}_p \, \partial / \partial Y_p
    }
  \right]
  {\mathbb{A}}_{\kappa}  
  \left< 
    {\Psi}_{Y} \! \left( \rn \right) 
  \right>
} & 
\nonumber \\ 
& = 
  \sum_{j_1 = 0}^{k_1}
  \left(
    \begin{array}{c}
      k_1 \\ j_1 
    \end{array}
  \right) \ldots 
  \sum_{j_P = 0}^{k_P}
  \left(
    \begin{array}{c}
      k_P \\ j_P 
    \end{array}
  \right)  
  \left[ 
    \dot{\prod_p}
    Y_p^{\underline{j_p}} \,  
  \right]
  {\mathbb{A}}_{\kappa}  
  \left< 
    {\Psi}_{Y + \left( k - j \right)} \! \left( \rn \right) 
  \right> 
\nonumber \\ 
& \equiv 
  \sum_{k^{\prime}}
  {\Lambda}_{k k^{\prime}} 
  {\Phi}_{k^{\prime}} . 
\label{eq:Glauber_moment_fact_prod}
\end{align}
The notation ${\dot{\prod}}_p$ stands for a product of terms over the
species index $p$ that is performed within each component of a row
vector over complexes $i$.  $Y_p \equiv \left[ y_p^i \right]$ is the
row vector of stoichiometric coefficients for species $p$ in each
complex, and both factorials $Y_p^{\underline{j_p}}$ and integer shift
operators $e^{ {\left( k - j \right)}_p \, \partial / \partial Y_p }$
are evaluated component-wise in indices $i$.  The resulting row vector
is contracted on the left with ${\mathbb{A}}_{\kappa}$.  The matrix
$\Lambda \equiv \left[ {\Lambda}_{k k^{\prime}} \right]$ is the
representation of the generator acting on the moment hierarchy.

\section{The eikonal approximation for the moment hierarchy}
\label{sec:eikonal_solns}

Eikonal approximations for the probability distribution
${\rho}_{\rn}$, in steady state, are based on a leading-exponential
expansion $ {\rho}_{\rn} \sim e^{- \Xi \left( \rn \right)}$ expressing
the large-deviation scaling. In Section \ref{subsec:LD scaling}, we
demonstrate how this approximation for the steady-state probability
distribution results in a corresponding approximation for the
factorial moments. We show first however that this approximation for
the factorial moments, in a large-number limit in steady state, may be
arrived at directly from the moment hierarchy as well.

The factorial moments Eq.~(\ref{eq:Glauber_moment_fact_prod}), under
the steady state condition, satisfy a hierarchy of relations $0 =
\sum_{k^{\prime}} {\Lambda}_{k k^{\prime}} {\Phi}_{k^{\prime}}$ By
dividing through by ${\Phi}_k$, each of these may be written as a
constraint on ratios of neighboring moments ${\Phi}_{k^{\prime}}$ to
${\Phi}_k$.  In one dimension (for CRN's involving only one species)
these become recursion relations, which may be solved by a method of
matched asymptotic
expansions~\cite{Krishnamurthy:CRN_moments:17,Smith:CRN_moments:17}
downward from $k \rightarrow \infty$ and upward from $k = 0$.

It will not be surprising that the steady-state condition is
equivalent to a condition $\mathcal{L} = 0$ in the Liouville
representation, a relation suggested by
Eq.~(\ref{eq:Liouville_eq}). We show in the following that the
steady-state condition for moments reduces in the large-system limit
to a particular form of the condition $\mathcal{L} = 0$ equivalent to
the coherent-state expression in the action~(\ref{eq:S_from_L_CS}).
However, as is well known in the literature on \textit{momentum-space
WKB methods}~\cite{Dykman:chem_paths:94,Assaf:mom_WKB:17}, the
$\mathcal{L} = 0$ submanifold contains ``momentum'' values (our
potential field $\eta$) consistent with derivatives of many different
distributions.  To isolate the stationary distribution requires
additional information from boundary conditions; which is the problem
that makes direct solution of higher-dimensional moment equations
difficult.  What is gained in passing to the continuum limit is a
non-local relation of the momentum values at each point through
stationary trajectories under variation of the
action~(\ref{eq:S_from_L_CS}) anchored in the
\emph{classical fixed points} of the rate equation, a result developed
extensively for the solution of boundary value and first-passage
problems by Freidlin and Wentzell~\cite{Freidlin:RPDS:98}, and since
generalized to many classes of stochastic
processes~\cite{Dykman:chem_paths:94,Assaf:mom_WKB:17}.

\subsection{Large-number limit of the generator in the moment
representation} 

The counterpart to a continuum limit in the moment hierarchy comes
from the observation that, if both particle numbers and the orders of
moments are large, the ratios of factorial moments at adjacent $k$
indices should change slowly with $k$.  If that is the case, then
factorial moments offset by finite lattice shifts will be related
through one ratio of moments in each direction $p$, raised to powers.
For the offset term appearing in
Eq.~(\ref{eq:Glauber_moment_fact_prod}), that limiting behavior
becomes
\begin{align}
  \left< 
    {\Psi}_{Y + \left( k - j \right)}^i \! \left( \rn \right) 
  \right> 
& \underset{n \gg 1}{\rightarrow}
  {
    \left. 
      \left(
        \prod_p
        R_p^{y^i_p - j_p}
      \right)
    \right|
  }_k
  \left< 
    \prod_p 
    {\rn}_p^{\underline{k_p}}
  \right> 
\nonumber \\ 
& = 
  {
    \left. 
      {\psi}_{Y-j}^i \! \left( R \right) 
    \right|
  }_k
  {\Phi}_k .
\label{eq:jacquard_walk}
\end{align}
Here ${\left.  R_p \right| }_k$ denotes $\left< {\rn}^{\underline{k +
1_p}} \right> / \left< {\rn}^{\underline{k}} \right>$: the ratio of
factorial moments incremented by the $p$th component of $k$, in a
neighborhood of the index $k$ of the reference moment ${\Phi}_k$.  $R
\equiv \left[ R_p \right]$ is the vector with these components at any
$k$.  The product of powers of $R_p$ in Eq.~(\ref{eq:jacquard_walk})
will be independent of the walk on the lattice between $k$ and $k + Y
- j$, and at large $k$ corrections from the exact matrices
${\Lambda}_{k k^{\prime}}$ will scale as $\mathcal{O} \! \left( 1 / k
\right)$ or smaller.

Dividing Eq.~(\ref{eq:Glauber_moment_fact_prod}) through by ${\Phi}_k$
and then taking the steady-state condition, produces the sum
\begin{align}
  0 
& = 
  \dot{\prod_p}
  \left[ 
    \sum_{j_p = 0}^{k_p}
    \left(
      \begin{array}{c}
        k_p \\ j_p 
      \end{array}
    \right)
    Y_p^{\underline{j_p}} \,  
    e^{
      - j_p \, \partial / \partial Y_p
    }
  \right]
  {\mathbb{A}}_{\kappa}
  {
    \left. 
      {\psi}_Y \! \left( R \right) 
    \right|
  }_k
\nonumber \\ 
& = 
  \sum_{j_1 = 0}^{k_1}
  \left(
    \begin{array}{c}
      k_1 \\ j_1 
    \end{array}
  \right) \ldots 
  \sum_{j_P = 0}^{k_P}
  \left(
    \begin{array}{c}
      k_P \\ j_P 
    \end{array}
  \right)  
  \left[ 
    \dot{\prod_p}
    Y_p^{\underline{j_p}} \,  
  \right]
  {\mathbb{A}}_{\kappa}  
  {
    \left. 
      {\psi}_{Y-j} \! \left( R \right) 
    \right|
  }_k
\nonumber \\ 
& = 
  \sum_{j_1 = 0}^{k_1}
  k_1^{\underline{j_1}}
  \ldots 
  \sum_{j_P = 0}^{k_P}
  k_P^{\underline{j_P}}
  \left[ 
    \dot{\prod_p}
    \left(
      \begin{array}{c}
        Y_p \\ j_p 
      \end{array}
    \right)
  \right] 
  e^{
    - j \cdot \partial / \partial Y
  } 
  {\mathbb{A}}_{\kappa}
  {
    \left. 
      {\psi}_Y \! \left( R \right) 
    \right|
  }_k
\nonumber \\ 
& = 
  \dot{\prod_p}
  \left[ 
    \sum_{j_p = 0}^{k_p}
    \left(
      \begin{array}{c}
        Y_p \\ j_p 
      \end{array}
    \right)
    k_p^{\underline{j_p}} 
    e^{
      - j_p \, \partial / \partial Y_p
    }
  \right]
  {\mathbb{A}}_{\kappa}
  {
    \left. 
      {\psi}_Y \! \left( R \right) 
    \right|
  }_k
\label{eq:Glauber_moment_SS_ctm}
\end{align}
The coefficient in the row vector on the third line of
Eq.~(\ref{eq:Glauber_moment_SS_ctm}) at index $i$ is the product of
binomial coefficients 
\begin{displaymath}
  \prod_p
  \left(
    \begin{array}{c}
      y^i_p \\ j_p 
    \end{array}
  \right)
\end{displaymath}
for $j_p \le y^i_p$ and zero otherwise.  If $k_p$ are large while
$j_p$ remain bounded by fixed $y^i_p$, up to corrections of order
$1/k_p$ we may approximate $k_p^{\underline{j_p}} \approx k_p^{j_p}$.
Then the sum on $j_p$ reduces back to
\begin{equation}
  0 = 
  {\psi}^T_Y \! 
  \left( 
    1 + k e^{- \partial / \partial Y}
  \right) 
  {\mathbb{A}}_{\kappa}
  {
    \left. 
      {\psi}_Y \! \left( R \right) 
    \right|
  }_k , 
\label{eq:Lamb_sum_ctm_lim}
\end{equation}
recognizable as the CRN Liouville operator from
Eq.~(\ref{eq:L_of_as}).  (As elsewhere, the argument $1 + k e^{-
\partial / \partial Y}$ is to be understood as a vector with
components $1 + k_p e^{- \partial / \partial Y_p}$ taking the place of
$a^{\dagger}_p$.)

The operator $e^{- \partial / \partial Y_p}$, acting on functions
${\psi}^i_Y \! \left( R \right)$ has the effect of reducing any such
product by one factor of $R_p$.  Thus we can rewrite the
expression~(\ref{eq:Lamb_sum_ctm_lim}) as
\begin{equation}
  0 = 
  {\psi}^T_Y \! 
  \left( 
    1 + k / R 
  \right) 
  {\mathbb{A}}_{\kappa}
  {
    \left. 
      {\psi}_Y \! \left( R \right) 
    \right|
  }_k .
\label{eq:L0_kR_args}
\end{equation}
Vanishing of $\Lambda \Phi$ in Eq.~(\ref{eq:jacquard_walk}) is thus
the same in the large-number limit as vanishing of the Liouville
operator in the steady-state condition~(\ref{eq:Liouville_eq}) for the
generating function, with numbers $R$ replacing the operator $a$ and
numbers $1 + k / R$ replacing the operator $a^{\dagger}$.

\subsection{Connection to the standard eikonal approximation}

\subsubsection{Preliminaries: stationary-path approximations in
Doi-Peliti theory, and the source of boundary conditions for the
steady state}

The Peliti construction~\cite{Peliti:PIBD:85,Peliti:AAZero:86}
likewise replaces $a^{\dagger}$ and $a$ in $\mathcal{L}$ with the
coherent-state parameters that are their eigenvalues. Stationary
trajectories arise in the saddle-point approximation of the functional
integral~(\ref{eq:DP_2FFI_CS}), where they are solutions to the
vanishing of the variational derivative of $S$. These stationary paths
are those which provide the eikonal approximation to the probability
density\footnote{For a discussion of when this property does or does
not apply to stationary paths used to calculate generating functions
and functionals, see~\cite{Smith:LDP_SEA:11}.  It will always apply to
the solutions computed here.  For much earlier work on a condition
equivalent to $\mathcal{L} = 0$ as a property of stationary
distributions, see~\cite{Eyink:action:96}.} and $\mathcal{L}
\! \left( {\xi}^{\dagger} , \xi \right) = 0$ is a property of these
paths.

The equations of motion from the
variation of $S$ in Eq.~(\ref{eq:S_from_L_CS}) are
\begin{align}
  \frac{d \xi}{d\tau}
& = 
  - \frac{\partial \mathcal{L}}{\partial {\xi}^{\dagger}}
& 
  \frac{d {\xi}^{\dagger}}{d\tau}
& = 
  \frac{\partial \mathcal{L}}{\partial \xi} . 
\label{eq:EOM_CS}
\end{align}
For the action~(\ref{eq:S_from_L_neta}) in number-potential fields,
the equivalent forms are:
\begin{align}
  \frac{d n}{d\tau}
& = 
  - \frac{\partial \mathcal{L}}{\partial \eta}
& 
  \frac{d\eta}{d\tau}
& = 
  \frac{\partial \mathcal{L}}{\partial n} . 
\label{eq:EOM_neta}
\end{align}

Solutions with ${\xi}^{\dagger} \equiv 1$ or $\eta \equiv 0$ reproduce
the mean-field rate equations~(\ref{eq:CRN_firstmom_EOM}) for $n$, which
then coincides with $\xi$. 

Potential solutions for $R$ joint with $1 + k / R$ come from points
along trajectories with ${\xi}^{\dagger} > 1$, describing improbable
excursions, for which $n \neq \xi$.  If the dimension of the subspace
of dynamical degrees of freedom in a CRN is some number $s \le P$, the
space of tangent vectors to stationary trajectories through a point
has dimension $2s$, and $\mathcal{L} = 0$ specifies a $\left( 2s-1
\right)$-dimensional manifold of possible solutions.

The saddle-point approximation (denoted $\sim$) to the value of the
generating function~(\ref{eq:DP_2FFI_neta}) is the integral over an
eikonal
\begin{align}
  {\phi}_T \! \left( z \right) 
& \sim
  {
    \left. 
      \left(
        \prod_p
        z_p^{{\rn}_p}
      \right)
      {\rho}_{\rn}
    \right|
  }_{\rn = {\bar{n}}_T}
  \sim 
  e^{
    \int_0^T d\tau
    \left( d_{\tau} \bar{\eta} \right)
    \bar{n} 
  } , 
\label{eq:MGF_LD_approx}
\end{align}
where the trajectory $\left( {\bar{\eta}}_{\tau}, {\bar{n}}_{\tau}
\right)$ satisfies equations~(\ref{eq:EOM_neta}) and has terminal
conditions for $\bar{\eta}$ derived from the variation of $n_T$, and
initial conditions for $\bar{n}$ derived from variation of ${\eta}_0$
including the contribution to the initial-value function ${\phi}_0 \!
\left( e^{{\eta}_0} \right)$.  

For the steady-state distribution, $T \rightarrow \infty$ and any
trajectory from the initial conditions passes arbitrarily close to one
of the attracting fixed points and remains there arbitrarily long
before escaping to the final values $\left( {\bar{\eta}}_T,
{\bar{n}}_T \right)$.  Therefore the relevant eikonals are the escape
trajectories from the attracting fixed points of the rate equation
~(\ref{eq:CRN_firstmom_EOM}), with initial values ${\eta}_0 = 0$ from
the converging solutions, and final values ${\bar{\eta}}_T$ determined
by $z$.

The distribution is then obtained from the Legendre transform of the
cumulant-generating function $\log {\phi}_T \! \left( z \right)$.
Choosing a value $z$ for which ${\bar{n}}_T = \rn$, and subtracting
${\bar{n}}_T \log z$, subtracts a total derivative of $\bar{\eta} \,
\bar{n}$ from the integral in Eq.~(\ref{eq:MGF_LD_approx}), leaving
\begin{align}
  {\rho}_{\rn} 
& \sim 
  e^{
    - \int^{\rn}
    \bar{\eta} \, d\bar{n}
  } 
  \equiv 
  e^{
    - \Xi \left( \rn \right) 
  } .
\label{eq:eff_pot_introd}
\end{align}
In Eq.~(\ref{eq:eff_pot_introd}) we introduce the \textit{effective
potential} $\Xi \left( \rn \right)$ defined as the eikonal integral
\begin{equation}
  \Xi \! \left( \rn \right) \equiv 
  \int^T d\tau \, \bar{\eta} \, d_{\tau} \bar{n} = 
  \int^{\rn} \bar{\eta} \, d\bar{n} .
\label{eq:eff_pot_def}
\end{equation}

\subsubsection{Large-deviations scaling and saddle-point
approximation of factorial moments}
\label{subsec:LD scaling}

The leading exponential dependence captured by the eikonal integral
gives the system's \textit{large-deviations}
scaling~\cite{Ellis:ELDSM:85} if, when we descale number $\rn$ by some
characteristic scale $n_C$\footnote{In the examples below, we will use
one of the fixed points of the first-moment
condition~(\ref{eq:CRN_firstmom_EOM}) as the scale factor, but that
choice is not essential.} as
\begin{equation}
  \frac{\rn}{n_C} \equiv x , 
\label{eq:descaled_number}
\end{equation}
and let $n_C$ become large, stationary properties such as trajectories
$\left( \bar{\eta} , \bar{n} / n_C \right)$ become asymptotically
independent of $n_C$.  In that case the effective potential separates
into scale and structure components~\cite{Touchette:large_dev:09} as 
\begin{align}
  \Xi \! \left( \rn \right) 
& = 
  n_C \int^{\rn} \bar{\eta} \, d\left( \bar{n} / n_C \right) \equiv 
  n_C \int^x \bar{\eta} \, d\bar{x}
  \equiv 
  n_C \hat{\Xi} \! \left( x \right) . 
\label{eq:descaled_eff_pot}
\end{align}
$\hat{\Xi} \! \left( x \right)$ is called the \textit{rate function}
of the large-deviations approximation for ${\rho}_{\rn}$.

From the large-deviations approximation to ${\rho}_{\rn}$, the
factorial moment at order $k$ has an implied approximation 
\begin{align}
  \left< {\rn}^{\underline{k}} \right> 
& \sim 
  \int dx n_C 
  e^{- n_C \hat{\Xi} \! \left( x \right)}
  \frac{
    \left( n_C x \right) ! 
  }{
    \left( n_C x - n_C y \right) !
  }
\nonumber \\
& \approx 
  n_C^k
  \int dx n_C 
  e^{- n_C \hat{\Xi} \left( x \right)}
  e^{
    n_C 
    \left\{
      x \left[ \log x - 1 \right] - 
      \left( x-y \right) 
      \left[
        \log \left( x-y \right) - 1
      \right]
    \right\}
  } , 
\label{eq:fact_mom_int_form}
\end{align}
where we have defined $y \equiv k / n_C$ in parallel with
Eq.~(\ref{eq:descaled_number}).  $\sim$ in the first line of
Eq.~(\ref{eq:fact_mom_int_form}) denotes the saddle-point
approximation to the effective potential, while $\approx$ in the
second line indicates the Stirling approximation for factorials.

The scale factor $n_C$ in Eq.~(\ref{eq:fact_mom_int_form}) separates
from a rate function including the contribution from the falling
factorials, which we therefore denote
\begin{align}
  \mathbf{\hat{\Xi}} \! \left( x ; y \right) 
& = 
  \hat{\Xi} \! \left( x \right) + 
  \left( x-y \right) 
  \left[
    \log \left( x-y \right) - 1
  \right] - 
  x \left[ \log x - 1 \right] 
\label{eq:0123_xy_eff_Pot}
\end{align}
A second saddle-point approximation in the integral yields an
expression for the descaled factorial moment as a finite sum 
\begin{equation}
  {\hat{\Phi}}_k \equiv 
  \frac{
    \left< {\rn}^{\underline{k}} \right> 
  }{
    n_C^k
  } \sim
  \sum_{\bar{x} \left( y \right)}
  e^{ - n_C
    \mathbf{\hat{\Xi}} \left( \bar{x} ; y \right) 
  } , 
\label{eq:fact_mom_SPA_form}
\end{equation}
where the saddle points $\bar{x} \! \left( y \right)$ are any
solutions to the equation
\begin{equation}
  \log 
  \left( 
    \frac{
      {\bar{x}}_p
    }{
      {\bar{x}}_p - y_p 
    } 
  \right) = 
  {
    \left. 
      \frac{d\hat{\Xi}}{dx_p}
    \right|
  }_{\bar{x}} = 
  {\eta}_p \! \left( \bar{x} \right) . 
\label{eq:fact_mom_saddle_pt}
\end{equation}

The ratio of two such solutions at values of $k$ separated by $1_p$
produces an expression for descaled moment ratios from
Eq.~(\ref{eq:jacquard_walk}) as
\begin{align}
\lefteqn{
  {\left. {\hat{R}}_p \right|}_y
  \equiv 
  \frac{
    {\left. R_p \right|}_k
  }{
    n_C
  } = 
  \frac{
    \left< {\rn}^{\underline{k+1_p}} \right> / 
    \left< {\rn}^{\underline{k}} \right> 
  }{
    n_C 
  }
} & 
\nonumber \\ 
& \underset{n_C \rightarrow \infty}{\rightarrow}
  \exp
  \left\{
    \frac{d}{dy_p}
    \mathbf{\hat{\Xi}} \! \left( \bar{x} ; y \right) 
  \right\}
\nonumber \\ 
& = 
  \exp
  \left\{
    \log \left( {\bar{x}}_p - y_p \right) + 
    \frac{ 
      {d\bar{x}}_p 
    }{ 
      dy_p
    } 
    {
      \left[
        \log 
        \left(
          \frac{x_p}{x_p-y_p}
        \right) - 
        \frac{d\hat{\Xi}}{dx_p}
      \right]
    }_{x = \bar{x}}
  \right\}
\nonumber \\ 
& = 
  {\bar{x}}_p - y_p .
\label{eq:hat_mom_rats}
\end{align}

From the saddle-point condition~(\ref{eq:fact_mom_saddle_pt}) and the
expression~(\ref{eq:hat_mom_rats}) for $\hat{R}$ it follows that 
\begin{align}
  y_p 
& = 
  {\bar{x}}_p 
  \left( 
    1 - e^{-{\eta}_p}
  \right)
\nonumber \\
  \frac{{\hat{R}}_p}{{\bar{x}}_p} 
& = 
  \frac{{\bar{x}}_p - y_p}{{\bar{x}}_p} = 
  e^{-{\eta}_p} . 
\label{eq:main_SP_relns}
\end{align}
Together with the definition of field
variables~(\ref{eq:xis_to_n_eta}) we confirm that ${\hat{R}}_p$
corresponds to the descaled coherent-state field ${\hat{\xi}}_p \equiv
{\xi}_p / n_C$.  Some algebra then gives 
\begin{align}
  1 + \frac{k_p}{R_p} 
& = 
  \frac{{\hat{R}}_p + y_p}{{\hat{R}}_p} = 
  \frac{{\bar{x}}_p}{{\hat{R}}_p} = 
  e^{{\eta}_p}
\label{eq:duplicate_main_SP_relns}
\end{align}
confirming that $1 + k / R$ corresponds to ${\xi}^{\dagger}$,
consistent with Eq. (\ref{eq:L0_kR_args}). Eq. (\ref{eq:hat_mom_rats})
and Eq. (\ref{eq:fact_mom_saddle_pt}) hence directly give us a
non-trivial approximation for the factorial moment ratios, consistent
with the eikonal approximation for the probability distribution.

\subsubsection{Direct derivation from the definitions of factorial
moments}

The mappings~(\ref{eq:main_SP_relns},\ref{eq:duplicate_main_SP_relns})
can be obtained directly from the definitions of factorial moments.  
The second line in Eq.~(\ref{eq:main_SP_relns}) is equivalent to
\begin{equation}
  \bar{\xi} = 
  \bar{n} - k . 
\label{eq:xi_n_k}
\end{equation}
The relations~(\ref{eq:Psi_psi_on_n_xi}) between actions of the
lowering operator give 
\begin{equation}
  {\bar{\xi}}_p = 
  \frac{
    \left< {\rn}^{\underline{k + 1_p}} \right>
  }{
    \left< {\rn}^{\underline{k}} \right>
  } . 
\label{eq:xi_as_rat}
\end{equation}
Then, from the definition~(\ref{eq:factorial_moment_not}),
\begin{equation}
  {\rn}^{\underline{k + 1_p}} = 
  {\rn}^{\underline{k}}
  \left( n_p - k_p \right) , 
\label{eq:n_und_k_rats}
\end{equation}
Eq.~(\ref{eq:xi_n_k}) is nothing more than the large-number-limiting
relation 
\begin{equation}
  \frac{
    \left< {\rn}^{\underline{k}} \rn \right>
  }{
    \left< {\rn}^{\underline{k}} \right>
  } = 
  \left< \rn \right> . 
\label{eq:large_n_exp}
\end{equation}  
In the background of a factorial moment, taking the expectation of an
additional factor of $n$ results in just the mean-field value to
$\mathcal{O} \! \left( n^0 \right)$.

The identification of Eq.~(\ref{eq:L0_kR_args})  as $\mathcal{L} \! 
\left( {\xi}^{\dagger} , \xi \right)$ along stationary trajectories is
thus confirmed, along with a map from $y = k / n_C$ to the position
$\bar{x}$ on the trajectory that controls the corresponding moment
ratio $\hat{R}$.  The mapping depends on convexity of the effective
potential $\Xi$.  If $\Xi$ is only locally convex, as occurs in
systems with multiple metastable fixed points, then $\hat{R}$ is
assigned to a stable branch by a rule equivalent to a coexistence
condition.  We provide examples in Sec.~\ref{sec:examples}.

\section{Examples}
\label{sec:examples}

A brief treatment will be given of three CRN examples, with complexity
added at each stage to illustrate the uses, variants, and limitations
of eikonal solutions.  The first example is a 1-species autocatalytic
system, illustrating the handling of bistability.  Methods are
introduced to characterize large-$n_C$ convergence of the eikonal
approximation and to estimate finite-size corrections.  The second
example is a symmetric 2-species cross-catalytic system that
reproduces the fixed-point structure of the 1-dimensional model along
its axis of symmetry, and introduces a boundary-value problem that is
difficult to solve without eikonal methods.  The third example is a
2-species CRN known as the Selkov model, which may be given a
fixed-point structure similar to the previous cases, but lacks axes of
symmetry.

Closed-form solution of the 1-dimensional CRN emphasizes why the
Hamilton-Jacobi dual coordinates $n$ and $\eta$ are also the index and
log-derivative of the stationary distribution, with $e^{\eta}$
representing a ratio of transport rates for probability current upward
and downward in $n$.  In higher dimensions where closed-form solutions
are no longer available, the stationarity condition under the
action~(\ref{eq:S_from_L_CS}) ensures that the planes normal to $\eta$
remain isoclines of the steady-state probability, so that the
interpretation of current balance remains as in the 1-dimensional
case.

In the symmetric cross-catalytic model,
values of the effective potential at the fixed points may be computed
directly from the recursion relations, without solving for eikonals.
In the final model without symmetry, eikonal methods are required to
estimate the relative residence times at fixed points, but they may
also exhibit caustics and other regions where the eikonal
approximation breaks down.

\subsection{Exactly solvable 1-variable autocatalytic CRN: multiple
fixed points and finite-size scaling}

\begin{figure}[ht]
  \begin{center} 
  \includegraphics[scale=0.6]{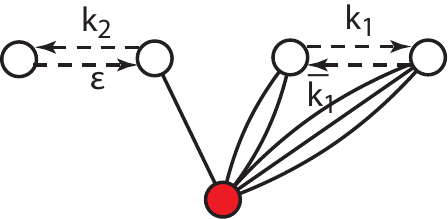}
  \caption{
  A network with one species and four complexes, which supports two
  metastable states at some values of the rate constants.  Filled
  circles are species; open circles are complexes.  Dashed lines
  between complexes indicate (directed) reactions; solid lines from
  species to complexes reflect the stoichiometric coefficients of the
  CRN.
    \label{fig:g_0123_unconn} 
  }
  \end{center}
\end{figure}

Fig.~\ref{fig:g_0123_unconn} shows a 1-species autocatalytic CRN,
which for suitable parameters possesses two metastable states,
expressed as two stable fixed points of the mean-field
equations~(\ref{eq:CRN_firstmom_EOM}).  The bipartite graphical
notation in the figure represents the decomposition of CRNs due to
Horn and Jackson~\cite{Horn:mass_action:72} and
Feinberg~\cite{Feinberg:def_01:87}, and is elaborated in detail
in~\cite{Krishnamurthy:CRN_moments:17,Smith:CRN_moments:17}.

The recursion relations for this model were considered in detail
in~\cite{Smith:CRN_moments:17}, where they were solved by methods of
matched asymptotic expansions and validated against Gillespie
simulations.  Here we use the model to illustrate the relation between
the moment and Liouville representations and to consider finite-size
deviations from the eikonal approximation.  Additional detail is given
in App.~\ref{sec:autocat_numerics}.

The reaction schema corresponding to Fig.~\ref{fig:g_0123_unconn} is 
\begin{align}
  \varnothing
& \xrightleftharpoons[k_2]{\epsilon}
  {\rm A} & 
  2 {\rm A} 
& \xrightleftharpoons[{\bar{k}}_1]{k_1}
  3 {\rm A} .
\label{eq:cubic_1spec_scheme}
\end{align}
where $A$ is the species name, and $\epsilon$, $k_2$, $k_1$, and
${\bar{k}}_i$ are reaction rate constants.  $\varnothing$ indicates
the \textit{null species}, which may stand for buffered external
species for which fluctuations are not considered, or depending on the
context, on \textit{de novo} creation or annihilation of particles.
The mean-field equation of motion~(\ref{eq:CRN_firstmom_EOM}) for the
number $n$ of $A$ particles in the system is
\begin{equation}
  \frac{dn}{d\tau} = 
  \epsilon - k_2 n + k_1 {n}^2 -{\bar{k}}_1 {n}^3 .
\label{eq:0123_y1A_rate_eqn}
\end{equation}

The exact recursion relation for $R_k$ following from
Eq. (\ref{eq:Glauber_moment_fact_prod}) was shown
in~\cite{Smith:CRN_moments:17} to be
\begin{widetext}
\begin{equation}
  R_k =  
  \frac{
      \epsilon + 
      \left( k-1 \right) \left( k-2 \right) k_1 
  }{
    k_2 - k_1 R_{k+1} +
    {\bar{k}}_1 R_{k+2} R_{k+1} - 
    \left( k-1 \right) 
    \left( 
      2k_1 - 2 {\bar{k}}_1 R_{k+1} 
    \right) + {\bar{k}}_1 
    \left( k-1 \right) \left( k-2 \right)
  } . 
\label{eq:rec_0123}  
\end{equation}
\end{widetext}
Descaled variables to extract the large-deviations system behavior
will be denoted
\begin{align}
  \epsilon 
& = 
  \varepsilon n_C 
& 
  k_2 
& = 
  {\kappa}_2
\nonumber \\
  k_1 
& = 
  {\kappa}_1 / n_C
& 
  {\bar{k}}_1 
& = 
  {\bar{\kappa}}_1 / n_C^2 
\nonumber \\ 
  n 
& \equiv 
  x n_C 
&
  R 
& \equiv 
  \hat{R} n_C 
\nonumber \\
  k 
& \equiv 
  y n_C
& 
  \mathcal{L} 
& = 
  \hat{\mathcal{L}} n_C .
\label{eq:0123_parm_descaling}
\end{align}
In terms of these Eq.~(\ref{eq:rec_0123}) has the large-$n_C$ and
large-$k$ limit
\begin{equation}
  \hat{R} = 
  \frac{
    \varepsilon + y^2 {\kappa}_1 
  }{
    {\kappa}_2 - 
    {\kappa}_1 \hat{R} + 
    {\bar{\kappa}}_1 {\hat{R}}^2 - 
    2 y 
    \left( 
      {\kappa}_1 - {\bar{\kappa}}_1 \hat{R}
    \right) + 
    {\bar{\kappa}}_1 y^2 
  } . 
\label{eq:0123_R_recur_from_CRNLP}
\end{equation}
Non-negative real solutions to 
\begin{equation}
  y = 
  \sqrt{
    \frac{
      {\kappa}_2 \hat{R} - \varepsilon
    }{
      {\kappa}_1 - {\bar{\kappa}}_1 \hat{R}
    }
  } - 
  \hat{R}
\label{eq:0123_two_y_sols}
\end{equation}
give the two branches of $\hat{R} \! \left( y \right)$ satisfying
Eq.~(\ref{eq:0123_R_recur_from_CRNLP}).

The Liouville function in number-potential fields, derived following
standard methods
(see~\cite{Smith:LDP_SEA:11,Krishnamurthy:CRN_moments:17,Smith:CRN_moments:17}),
is
\begin{align}
  \hat{\mathcal{L}} 
& = 
  \left( 1 - e^{\eta} \right)
  \left( 
    \varepsilon + {\kappa}_1 x^2
  \right) + 
  \left( 1 - e^{- \eta} \right)
  \left( 
    {\kappa}_2 + {\bar{\kappa}}_1 x^2
  \right) x . 
\label{eq:0123_L_form}
\end{align}
It is immediate to check that under the
assignments~(\ref{eq:main_SP_relns}),
Eq.~(\ref{eq:0123_R_recur_from_CRNLP}) re-arranges to
\begin{align}
  0 
& = 
  -\varepsilon + 
  {\kappa}_2 \hat{R} - 
  {
    \left( y + \hat{R} \right)
  }^2
  \left( 
    {\kappa}_1 - {\bar{\kappa}}_1 \hat{R}
  \right) = 
  \frac{\hat{R}}{y}
  \hat{\mathcal{L}} . 
\label{eq:onespec_yR_sol}
\end{align}

For a 1-dimensional system, the condition $\mathcal{L} = 0$ is
sufficient to solve jointly for $x$ and $\eta$, giving a relation
\begin{equation}
  \hat{R} = 
  \frac{
    \varepsilon + {\kappa}_1 x^2
  }{
    {\kappa}_2 + {\bar{\kappa}}_1 x^2
  } 
\label{eq:twospec_diag_ray_sol}
\end{equation}
that is equivalent under the mapping~(\ref{eq:main_SP_relns}) to
Eq.~(\ref{eq:0123_two_y_sols}).  It follows also that escape
trajectories are the time reverses of the solutions to the
first-moment equations~(\ref{eq:CRN_firstmom_EOM}).  

These properties of the continuum limit express the more basic
property that in 1-dimensional systems at steady states, upgoing and
downgoing probability flows must balance through each position.  If
only steps $\delta \rn$ of size 1 are allowed, as is the case for this
CRN, the process is called a \textit{birth-death} process, and
probability flux balance renders its steady-state probability
distribution solvable in closed
form~\cite{vanKampen:Stoch_Proc:07}.\footnote{Balance of upgoing and
downgoing probability currents between any two adjacent indices
$\left( \rn , \rn + 1 \right)$ is written in terms of the descaled
rate constants as
\begin{equation}
  \frac{
    {\rho}_{\rn + 1} 
  }{
    {\rho}_{\rn}
  } = 
  \frac{n_C}{\left( \rn + 1 \right)}
  \left( 
  \frac{
    \varepsilon + {\kappa}_1
    \rn \left( \rn - 1 \right) / n_C^2
  }{
    {\kappa}_2 + {\bar{\kappa}}_1
    \rn \left( \rn - 1 \right) / n_C^2
  }
  \right) 
\label{eq:birth_death_exact}
\end{equation}
Eq.~(\ref{eq:birth_death_exact}) is the exact evaluation of $\hat{R} /
x = e^{-\eta}$ and goes at large $n_C$ to $\hat{R}$ in
Eq.~(\ref{eq:twospec_diag_ray_sol}).  $\log
\left( {\rho}_{\rn} \right)$ is then the (exact) discrete sum
approximated by the continuum integral~(\ref{eq:eff_pot_introd}).
\label{fn:1D_birth_death}}

\subsubsection*{Finite-size scaling of the transition regions using
the eikonal approximation to the effective potential}

Ordinary position-space WKB methods can be used 
at finite $n_C$ to estimate low-order
moments.  Of particular interest for multistable systems is the order
at which fluctuations transition between control by different fixed
points.  To obtain an estimate, we make a simplifying approximation of
the integral~(\ref{eq:fact_mom_int_form}) as a sum of Poisson
distributions at the stable fixed points with weights from the
effective potential
\begin{equation}
  {\rho}_x \approx 
  \frac{
    e^{- \Xi \left( {\bar{x}}_1 \right)}
    \delta \! \left( x - {\bar{x}}_1 \right) + 
    e^{- \Xi \left( {\bar{x}}_3 \right)}
    \delta \! \left( x - {\bar{x}}_3 \right)
  }{
    e^{- \Xi \left( {\bar{x}}_1 \right)} + 
    e^{- \Xi \left( {\bar{x}}_3 \right)}
  } . 
\label{eq:0123_WKB_dist}
\end{equation}
The $k$th-order moment is approximated, without making use of the
curvature of the effective potential in neighborhoods of the fixed
points, as 
\begin{equation}
  \left<
    {\rn}^{\underline{k}}
  \right> \approx 
  \frac{
    e^{- \Xi \left( {\bar{x}}_1 \right)}
    {\bar{x}}_1^k + 
    e^{- \Xi \left( {\bar{x}}_3 \right)}
    {\bar{x}}_3^k 
  }{
    e^{- \Xi \left( {\bar{x}}_1 \right)} + 
    e^{- \Xi \left( {\bar{x}}_3 \right)}
  }
\label{eq:0123_WKB_fact_moms}
\end{equation}

The dominant term in the ratio~(\ref{eq:0123_WKB_fact_moms}) shifts
from the lower to the higher fixed point where the weighted effective
potentials are equal, corresponding to an equal-area rule under the
curve for the ``force'' in the two-variable effective potential 
\begin{equation}
  \frac{\partial}{\partial \bar{x}}
  \mathbf{\Xi} \! \left( \bar{x} ; y \right) = 
  \eta \! \left( \bar{x} \right) - 
  \log 
  \left(
    \frac{\bar{x}}{\bar{x} - y}
  \right) , 
\label{eq:0123_xy_spinodal}
\end{equation}
as in the coexistence condition for systems with first-order phase
transitions.  The crossover point in this approximation is 
\begin{equation}
  y_0 = 
  \frac{
    \hat{\Xi} \! \left( {\bar{x}}_3 \right) - 
    \hat{\Xi} \! \left( {\bar{x}}_1 \right)
  }{
    \log {\bar{x}}_3 - 
    \log {\bar{x}}_1 
  } .
\label{eq:0123_y0_solve}
\end{equation}

\begin{figure}[ht]
\begin{center} 
  \includegraphics[scale=0.4]{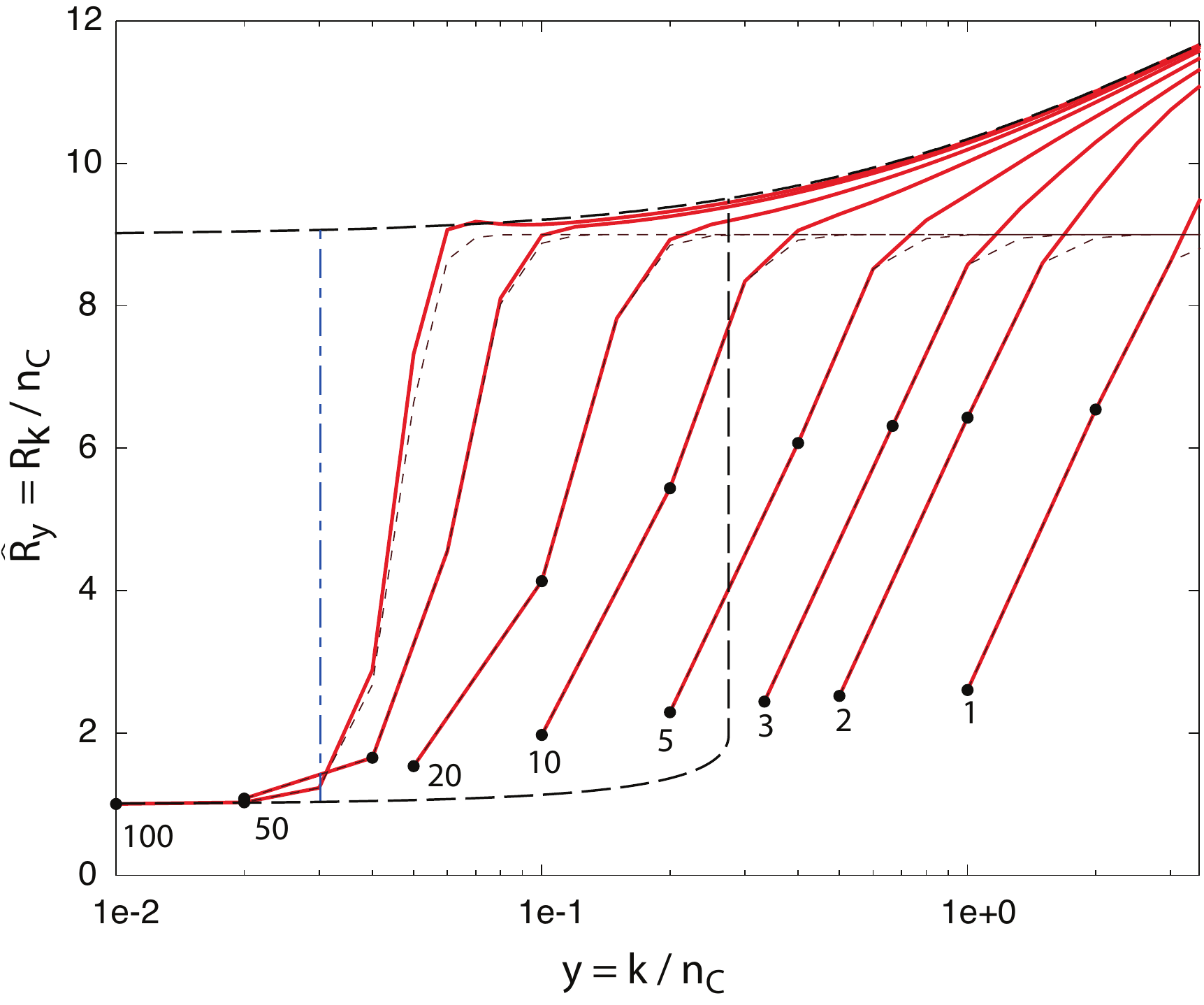} 
  \caption{
  Finite-$n_C$ evaluations (solid) of the exact recursion
  relation~(\ref{eq:rec_0123}), initiated at ${\hat{R}}_0$ and
  ${\hat{R}}_1$ (circles) 
  using values from Eq.~(\ref{eq:0123_WKB_fact_moms}).  That
  approximation (red-dashed) saturates at ${\bar{x}}_3 /
  {\bar{x}}_1$.  The lower and upper branches of the solution to
  the quadratic equation~(\ref{eq:0123_R_recur_from_CRNLP}) are shown
  in black-dash (vertical connection is the upper limit where
  solutions exist on the lower branch).  Equal-area value $y_0$ of
  Eq.~(\ref{eq:0123_y0_solve}), shown connecting the upper and lower
  branches $R \! \left( y \right)$ is approached by the recursion
  solutions at large $n_C$.
  \label{fig:scaled_moms_bistable_good_ks} 
  } 
\end{center}
\end{figure}

Fig.~\ref{fig:scaled_moms_bistable_good_ks} compares the
approximation~(\ref{eq:0123_WKB_fact_moms}) to the exact
recursion~(\ref{eq:rec_0123}) at finite $n_C$, and characterizes the
convergence to the eikonal approximation at large $n_C$.  When the
small-$k$ asymptotic expansion is initialized with values
${\hat{R}}_0$ and ${\hat{R}}_1$ derived from
Eq.~(\ref{eq:0123_WKB_dist}), the exact recursion solutions remain
close to that approximation over the interval between ${\bar{x}}_1$
and ${\bar{x}}_3$.  The WKB solution has captured
much of the finite-size scaling even though the effective potential is
approximated with the large-deviations scaling limit.  When $y$ is
large enough that $\hat{R}$ exceeds the upper fixed-point value
${\bar{x}}_3$, the recursion solutions converge on the exact continuum
value~(\ref{eq:0123_two_y_sols}).

\subsection{Two-dimensional cross-catalytic CRN, and eikonal
approximation at boundaries}

\begin{figure}[ht]
  \begin{center} 
  \includegraphics[scale=0.6]{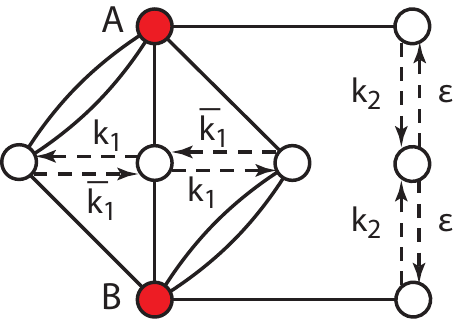}
  \caption{
  A network with two species labeled $A$ and $B$ in a cross-catalytic
  configuration.  The fixed points for both $A$ and $B$ are the same
  as those for $A$ in the CRN of Fig.~\ref{fig:g_0123_unconn}.
    \label{fig:two_species_bistable} 
  }
  \end{center}
\end{figure}

To demonstrate the use of eikonals in a model with similar qualitative
behavior that is not solvable in closed form,
Fig.~\ref{fig:two_species_bistable} shows a symmetric cross-catalytic
CRN with the same fixed point structure along its axis of symmetry as
the previous CRN.  Recursion relations in this model were solved and
compared to numerical simulations
in~\cite{Krishnamurthy:CRN_moments:17,Smith:CRN_moments:17} only along
the axis of symmetry.

The reaction schema corresponding to
Fig.~\ref{fig:two_species_bistable} is 
\begin{align}
  \varnothing
& \xrightleftharpoons[k_2]{\epsilon}
  {\rm A} & 
  {\rm A} + {\rm B} 
& \xrightleftharpoons[{\bar{k}}_1]{k_1}
  2 {\rm A} + {\rm B} 
\nonumber \\ 
  \varnothing
& \xrightleftharpoons[k_2]{\epsilon}
  {\rm B} & 
  {\rm B} + {\rm A} 
& \xrightleftharpoons[{\bar{k}}_1]{k_1}
  2 {\rm B} + {\rm A} , 
\label{eq:cubic_2spec_scheme}
\end{align}
and its mean-field equation of motion~(\ref{eq:CRN_firstmom_EOM}) for
numbers $n_A$ of species $A$ and $n_B$ of species $B$ is
\begin{align}
  \frac{d n_A}{d\tau}
& = 
  \epsilon - 
  k_2 n_A + 
  n_B 
  \left( 
    k_1 n_A - 
    {\bar{k}}_1 n_A^2 
  \right) 
\nonumber \\
  \frac{d n_B}{d\tau}
& = 
  \epsilon - 
  k_2 n_B + 
  n_A 
  \left( 
    k_1 n_B - 
    {\bar{k}}_1 n_B^2 
  \right) . 
\label{eq:rate_ab_cross}
\end{align}
Examples of the mean-field flow solutions, and more detailed
properties of the escape eikonals, are provided in
App.~\ref{sec:crosscat_numerics}, and we summarize only novel features
here.

\begin{figure}[ht]
\begin{center} 
  \includegraphics[scale=0.45]{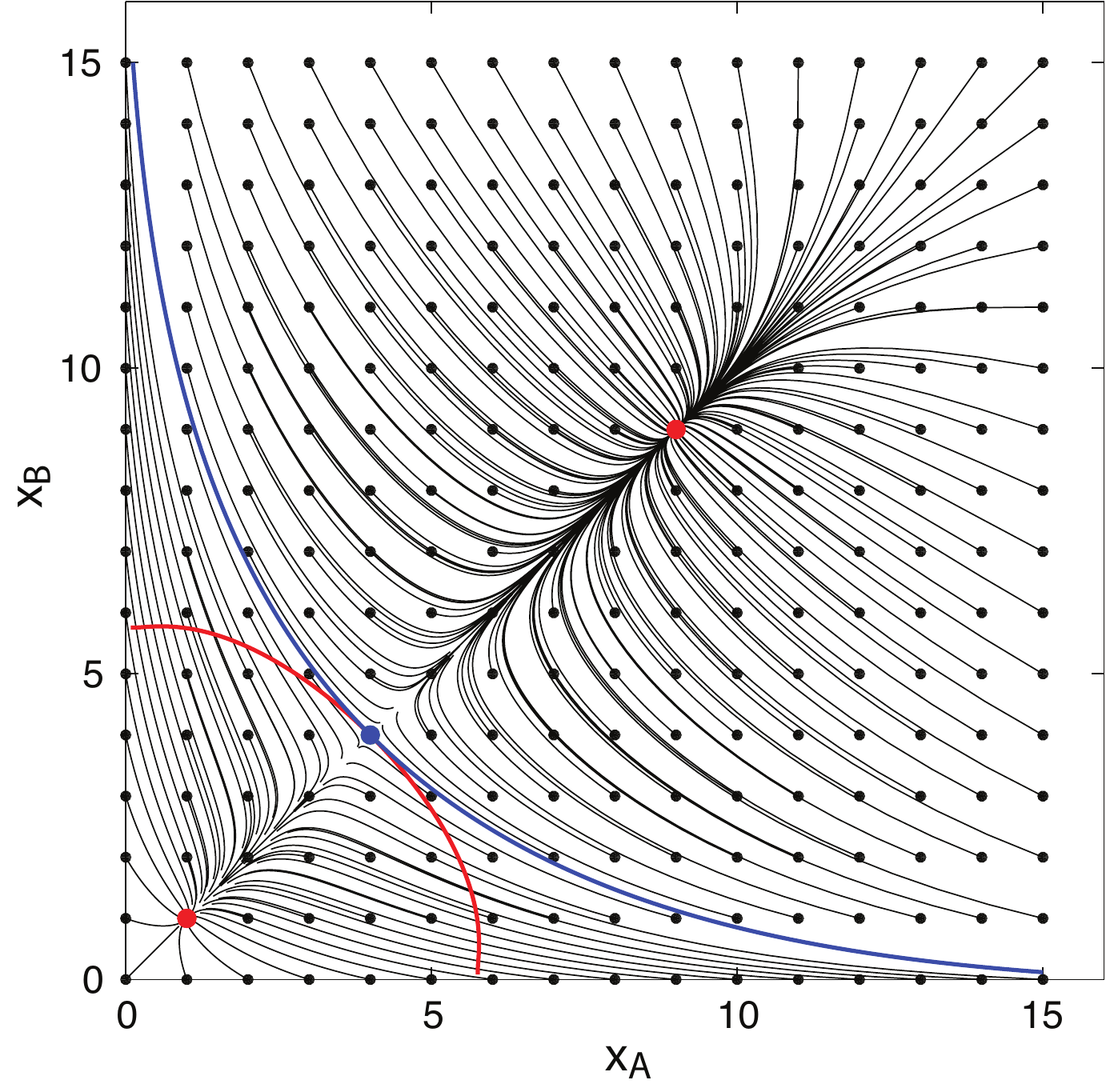} 
  \caption{
  Skeleton of trajectories determining the topological domains of the
  CRN from Fig.~\ref{fig:two_species_bistable}.  Solutions to the
  mean-field equations~(\ref{eq:rate_ab_cross}) from
  Fig.~\ref{fig:two_spec_ctm_class_flows} in the appendix are shown in
  thin black.  A converging separatrix (blue heavy) passing through
  the saddle separates the basins of attraction of the two fixed
  points.  An escape separatrix (red heavy), the locus of points
  reached with equal probability from either fixed point, also passes
  through the saddle and is otherwise contained in the basin of
  attraction of the lower fixed point.
  \label{fig:two_spec_ctm_escape_sep_bkgnd} 
  } 
\end{center}
\end{figure}

The topology of the space of trajectories is delimited by the skeleton
shown in Fig.~\ref{fig:two_spec_ctm_escape_sep_bkgnd}.  An attracting
separatrix divides the basins of attraction of the two stable fixed
points.  For this CRN (see App.~\ref{sec:crosscat_numerics}), escape
eikonals never cross, so they form a foliation of the space of $n$
values.  They are partitioned across the axis of symmetry by escape
eikonals from the stable fixed points either to the saddle or to $x_A
= x_B = 0$ or $\infty$, along which the integral $\int \eta \, dx$
gives the relative depth of the minima of the effective potential
$\hat{\Xi} \! \left( x \right)$.  An escape separatrix, the locus of
points with equal probability to be reached upon escape from either
fixed point, emanates from the saddle, and is entirely contained
within the basin of attraction of the lower fixed point.

\subsubsection*{Eikonal solutions along the non-asymptotic boundary}

The $A \leftrightarrow B$ exchange symmetry of this model was used
in~\cite{Krishnamurthy:CRN_moments:17,Smith:CRN_moments:17} to reduce
the recursion relations for moment ratios to a pair of self-consistent
solutions similar in form to those for the 1-species autocatalytic
CRN.  The generator in the moment
representation~(\ref{eq:Glauber_moment_fact_prod}) is a 2-dimensional
discrete Laplacian, given in Eq.~(\ref{eq:twospec_rull_gen}).  Its
solutions off the diagonal depend on boundary conditions at $k_A = 0$
or $k_B = 0$ where the nonzero moment order need not lie in any
asymptotic regime.  In the limit $y_B \rightarrow 0$, an estimate for
$y_A$ is given by
\begin{equation}
  y_A = 
  \frac{1}{{\hat{R}}_B}
  \left( 
    \frac{
      {\kappa}_2 {\hat{R}}_A - \varepsilon
    }{
      {\kappa}_1 - {\bar{\kappa}}_1 {\hat{R}}_A
    }
  \right) - 
  {\hat{R}}_A
\label{eq:twospec_yb_bdry_sol}
\end{equation}
In Eq.~(\ref{eq:twospec_yb_bdry_sol}) ${\hat{R}}_B \! \left( y_A , 0
\right)$ is not given, but for $y \sim 1$ where the boundary is not
very far from the diagonal, the approximation ${\hat{R}}_B \approx
{\hat{R}}_A$ permits a closed-form comparison with
Eq.~(\ref{eq:0123_two_y_sols}) which applies to both $y_A$ and $y_B$
along the diagonal.  An approximation of the effective potential by
time-reversing trajectories, developed in
App.~\ref{sec:crosscat_numerics} shows that the transfer of control
between the two fixed points occurs along a roughly linear contour
$y_A + y_B \approx 2 y_0$ from Eq.~(\ref{eq:0123_y0_solve}).

\begin{figure}[ht]
\begin{center} 
  \includegraphics[scale=0.45]{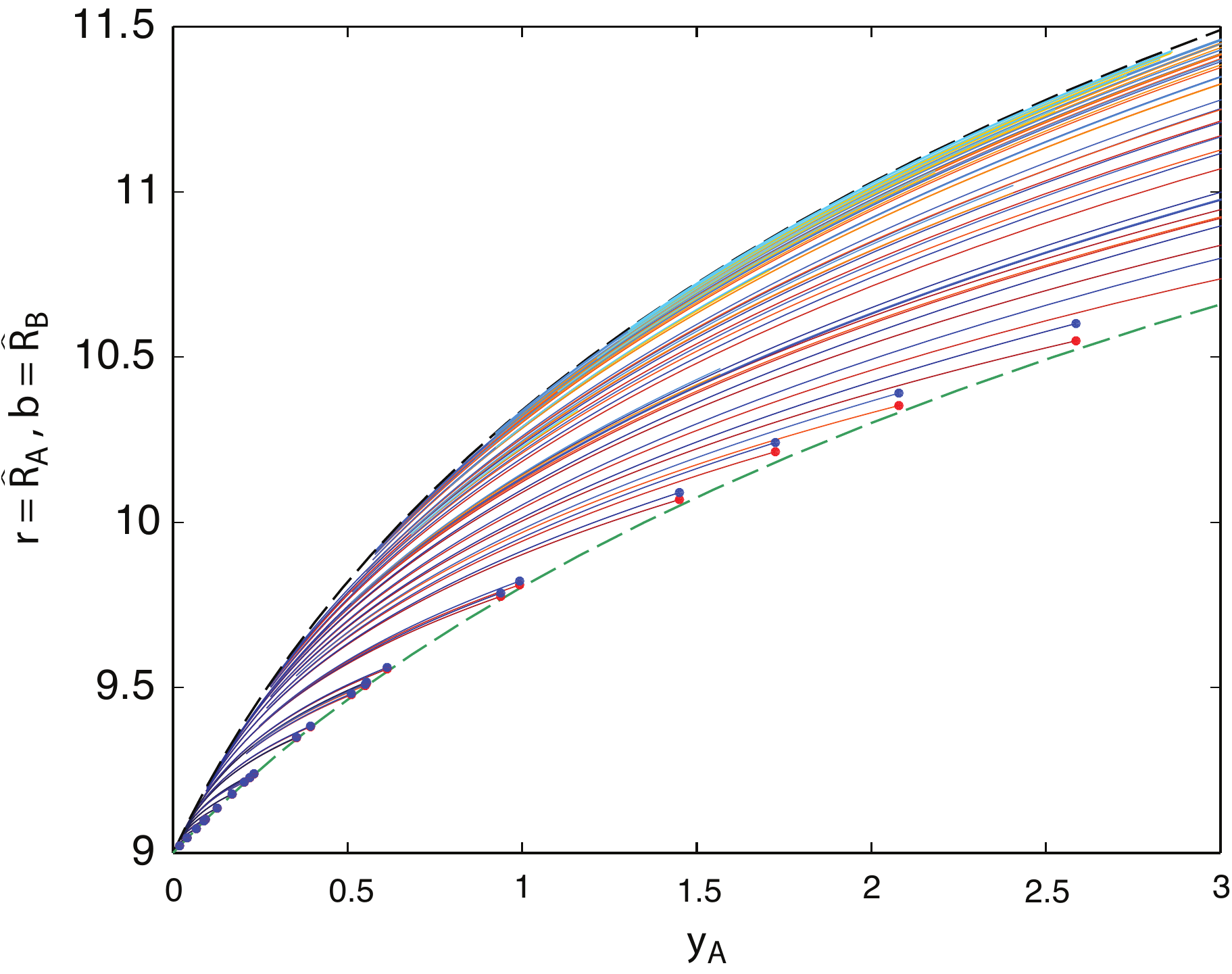} 
  \caption{
  ${\hat{R}}_A \! \left( y_A, y_B \right)$ (red-toned lines and
  symbols) and ${\hat{R}}_B \! \left( y_A, y_B \right)$ (blue-toned
  lines and symbols) for eikonals emanating from neighborhoods of the
  escape trajectory between the lower fixed point and the saddle into
  the region $y_B < y_A$.  Trajectories terminating in dots are
  truncated at the last point along the discrete numerical integral
  before the corresponding $y_B$ passes below zero.  Upper black
  dashed curve is the eikonal solution~(\ref{eq:0123_two_y_sols})
  along the diagonal, and lower green dashed curve is the
  estimation~(\ref{eq:twospec_yb_bdry_sol}).  The asymmetry
  ${\hat{R}}_B - {\hat{R}}_A \ge 0$ remains small over the regions
  shown and increases with increasing $x_A, x_B \lesssim
  {\bar{x}}_3$. 
  \label{fig:two_spec_ray_bdry_diag_comp_limits} 
  } 
\end{center}
\end{figure}

Fig.~\ref{fig:two_spec_ray_bdry_diag_comp_limits} illustrates the use
of eikonals that emerge from neighborhoods of the diagonal to
approximate the boundary conditions at $k_B = 0$.  It emphasizes that
the boundary values at any point are governed not by propagation along
the boundary, but by the values of moment ratios near the
diagonal. 
Thus, in a population process with $P$ fluctuating species,
although the boundary with undetermined conditions is a $\left( 2P - 1
\right)$-dimensional hypersurface, the behavior along the boundary is
controlled by neighborhoods of a discrete set of interior fixed points
and the parameters of $\mathcal{L}$.  

\subsection{Selkov model: asymmetric autocatalysis with breakdowns in
the simple eikonal approximation} 

\begin{figure}[ht]
  \begin{center} 
  \includegraphics[scale=0.6]{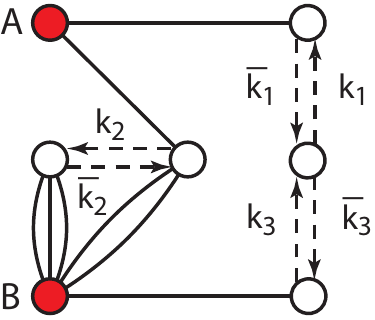}
  \caption{
  The Selkov model: an asymmetric autocatalytic CRN with two species
  $A$ and $B$.  The model can yield fixed points with the same spacing
  and similar effective potential depths to the cross-catalytic
  network of Fig.~\ref{fig:two_species_bistable}, but on a line of
  negative rather than positive slope.
    \label{fig:g_selkov} 
  }
  \end{center}
\end{figure}

To present a case in which even the coarsest WKB
approximation~(\ref{eq:0123_WKB_dist}) cannot readily be obtained
without continuum methods, Fig.~\ref{fig:g_selkov} shows a CRN known
as the Selkov model, which was treated with eikonal methods by Dykman
\textit{et al.}~\cite{Dykman:chem_paths:94}.  The Selkov model is an
asymmetric autocatalytic system for which we can choose fixed points
and residence times similar to those in the cross-catalytic CRN,
though the eikonals in the two models show qualitative differences.

\begin{figure}[ht]
\begin{center} 
  \includegraphics[scale=0.45]{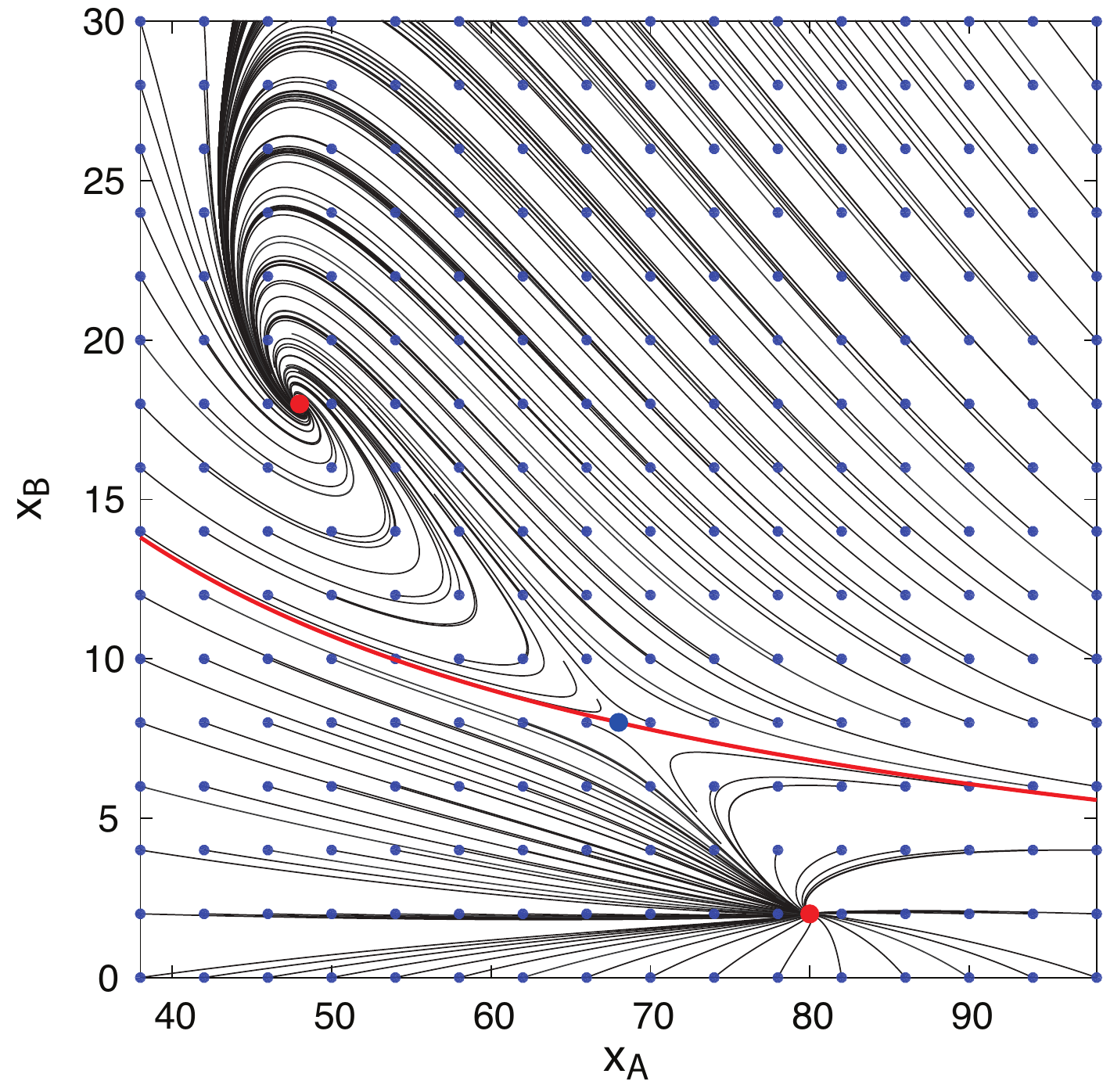} 
  \caption{
  Flowfields from the mean-field equations~(\ref{eq:CRN_firstmom_EOM})
  for the Selkov model.  Parameters were chosen to place fixed points
  at $\left( {\bar{x}}_A, {\bar{x}}_B \right) = \left\{ \left( 2, 80
  \right) , \left( 8, 68 \right) , \left( 18, 48 \right) \right\}$.
  Trajectories satisfy $d_{\tau} \left( {\rn}_{\rm A} + {\rn}_{\rm B}
  \right) = 0$ where they pass through the
  line~(\ref{eq:Selkov_linpart_solve}) connecting the fixed points.
  \label{fig:selkov_ctm_class_flows} 
  } 
\end{center}
\end{figure}

\begin{figure}[ht]
\begin{center} 
  \includegraphics[scale=0.45]{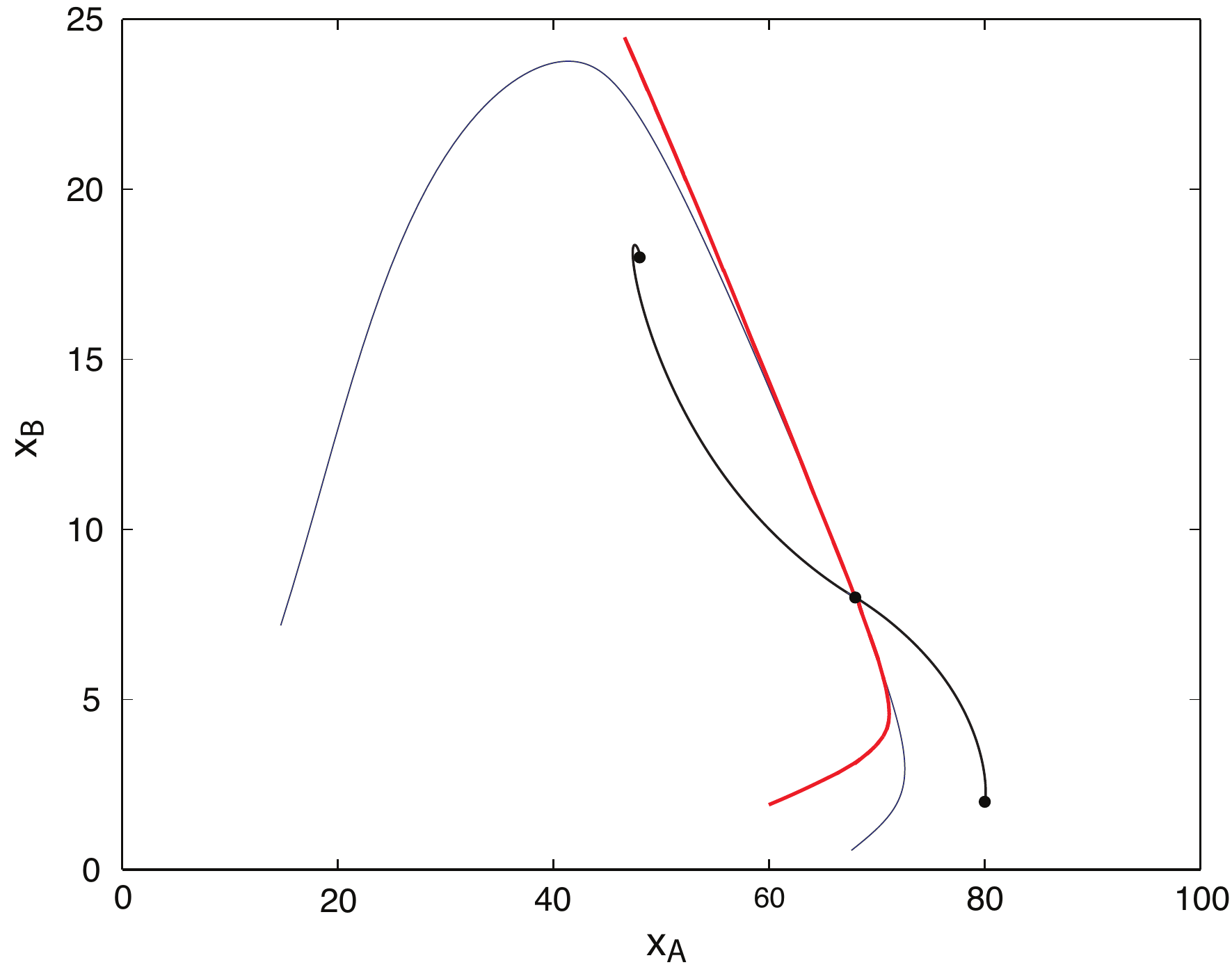} 
  \caption{
  Skeleton for escape trajectories in the Selkov model.  Eikonals from
  stable fixed points to the saddle are (heavy black).  Trajectories
  originating counterclockwise from those double back and cross,
  creating a singular measure, in caustics (heavy red).  The innermost
  limit of these rebounding trajectories is the limiting ray (thin
  black).  Each point between the limiting ray and caustic is crossed
  by two trajectories, and each trajectory is crossed continuously
  before that point, requiring modification of the eikonal
  approximation.
  \label{fig:selkov_eiks_limrays} 
  } 
\end{center}
\end{figure}

The reaction schema for Fig.~\ref{fig:g_selkov} is
\begin{align}
  \varnothing
& \xrightleftharpoons[{\bar{k}}_1]{k_1}
  {\rm A} & 
  {\rm A} + 
  2 {\rm B} 
& \xrightleftharpoons[{\bar{k}}_2]{k_2}
  3 {\rm B} & 
  {\rm B} 
& \xrightleftharpoons[{\bar{k}}_3]{k_3}
  \varnothing . 
\label{eq:Selkov_scheme}
\end{align}
The fixed points of the Selkov model lie along a line
\begin{align}
  n_{\rm A} + 
  n_{\rm B} = 
  \frac{
    2 \left( k_1 + {\bar{k}}_3 \right)
  }{
    \left( {\bar{k}}_1 + k_3 \right)
  } - 
  \frac{
    \left( {\bar{k}}_1 - k_3 \right) 
  }{
    \left( {\bar{k}}_1 + k_3 \right)
  }
  \left( 
    n_{\rm A} - 
    n_{\rm B}  
  \right) , 
\label{eq:Selkov_linpart_solve}
\end{align}
Fig.~\ref{fig:selkov_ctm_class_flows} shows the solutions to the
mean-field equations~(\ref{eq:CRN_firstmom_EOM}) for $\left(
{\bar{k}}_1 - k_3 \right) / \left( {\bar{k}}_1 + k_3 \right) = - 1/3$.  
Fig.~\ref{fig:selkov_eiks_limrays} shows the skeleton eikonals for the
escape system of the Selkov model, reproducing features observed
in~\cite{Dykman:chem_paths:94}.\footnote{We do not try here to
identify the escape separatrix, which in~\cite{Dykman:chem_paths:94}
falls within the regions between the caustics and limiting rays, areas
where each state is reached by multiple trajectories.} Further detail
is provided in App.~\ref{sec:selkov_numerics}.

In contrast to the cross-catalytic CRN~(\ref{eq:cubic_2spec_scheme}),
in which every point $n$ is traversed by a unique escape trajectory
and has a well-defined eikonal approximation to the escape
probability, the Selkov model at parameters shown possesses two
caustics emanating from the saddle point.  Caustics are regions where 
the momentum-space WKB approximation breaks down
as stationary trajectories from
finite measure converge on a subsurface of measure zero.  These
caustics are formed as escapes diverging counterclockwise from the
eikonal to the saddle turn back and cross.  The innermost returning
ray on each side of the saddle is the limiting ray, the boundary of
the domain of validity of the simple stationary-path approximation.

\begin{figure}[ht]
\begin{center} 
  \includegraphics[scale=0.35]{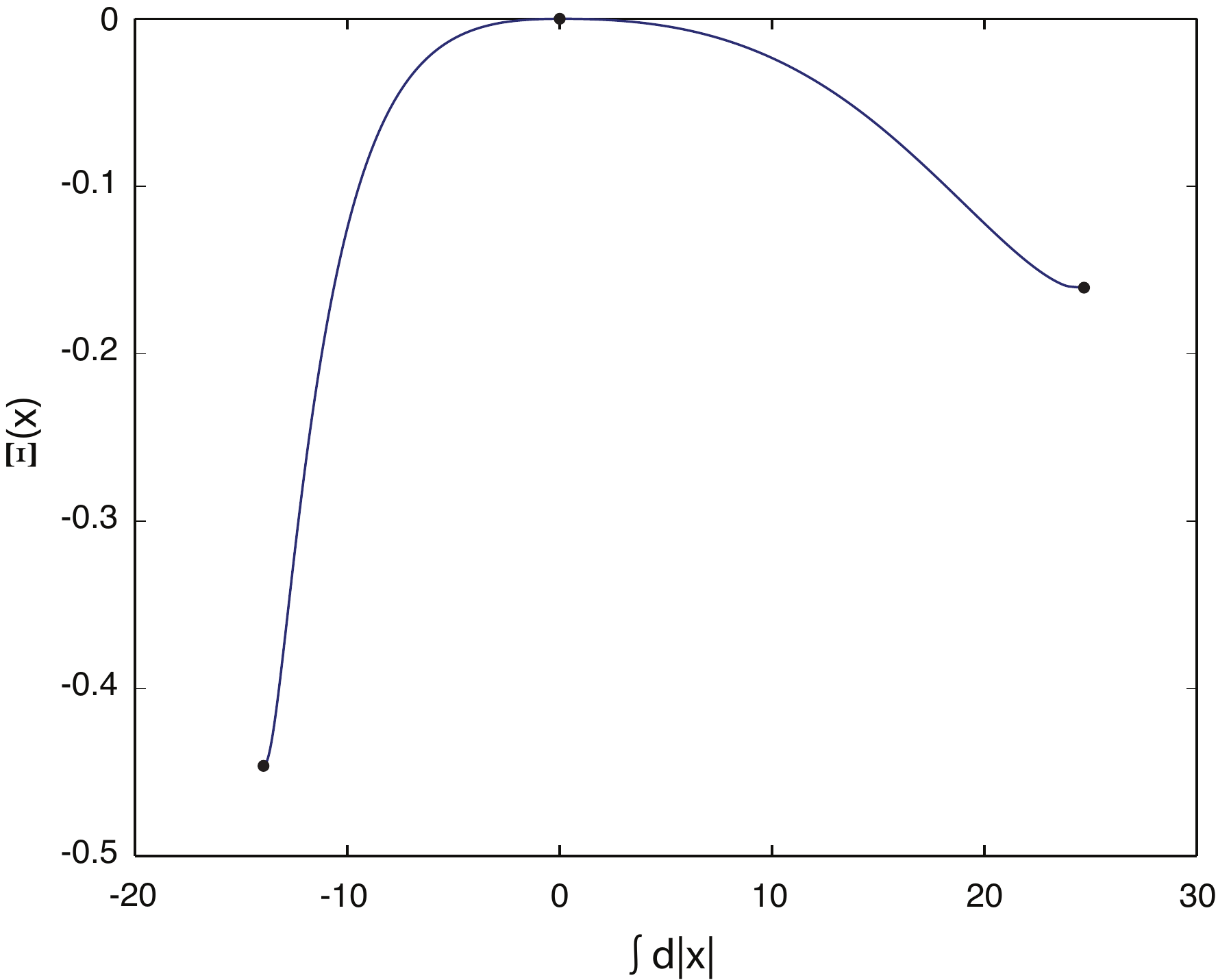} 
  \caption{
  $\int \eta \, dn$ along the eikonals from the two stable fixed
  points to the saddle gives the well depths of the effective
  potential.  For these parameters $\hat{\Xi} \! \left( {\bar{x}}_1
  \right) = - 0.4462$ and $\hat{\Xi} \! \left( {\bar{x}}_3 \right) = -
  0.1606$.  The difference, $0.2856$, is $\sim 4 \times$ the
  difference for the 1-dimensional autocatalytic model shown in Fig.~7
  of~\cite{Smith:CRN_moments:17}, and about twice the difference of
  the cross-catalytic model shown in
  Fig.~\ref{fig:two_spec_pot_rev_flat}.
  \label{fig:selkov_SaddleEiks_eff_pot} 
  } 
\end{center}
\end{figure}

Finite-size effects may again be estimated by the coarse WKB
approximation~(\ref{eq:0123_WKB_dist}), but now the escape equations
of motion must be solved to estimate the relative depths of the
effective potential at the fixed point.
Fig.~\ref{fig:selkov_SaddleEiks_eff_pot} shows $\int \eta \, dn$ along
the escape eikonals from the two stable fixed points to the saddle in
Fig.~\ref{fig:selkov_eiks_limrays}.  These give values $\hat{\Xi}
\! \left( {\bar{x}}_1 \right)$ and $\hat{\Xi} \! \left( {\bar{x}}_3
\right)$ determining relative residence times and the first-passage
time as a function of $n_C$.

Using the convention for large-deviations scaling that $\bar{n} = n_C
\bar{x}$ with fixed points 
\begin{align}
  {\bar{x}}_1
& = 
  \left(
    80 , 2 
  \right) , 
& 
  {\bar{x}}_3
& = 
  \left(
    48 , 18
  \right) , 
\label{eq:selkov_scaled_FP}
\end{align}
control of both moment ratios ${\hat{R}}_A$ and ${\hat{R}}_B$ is
expected to switch between the fixed points along the line
\begin{align}
  \begin{array}{c}
    \left[ 
      \begin{array}{cc}
        y_A & y_B
      \end{array}
    \right] \\
    \phantom{\mbox{}}
  \end{array}
  \left[ 
    \begin{array}{c}
      \log \left( 48 / 80 \right) \\
      \log \left( 18 / 2 \right) 
    \end{array}
  \right] 
& = 
  \hat{\Xi} \! \left( {\bar{x}}_3 \right)  - 
  \hat{\Xi} \! \left( {\bar{x}}_1 \right) , 
\label{eq:selkov_twomin_cross_short}
\end{align}
similar to the linear front $y_A + y_B = 2 y_0$ seen for the
cross-catalytic CRN~(\ref{eq:cubic_2spec_scheme}).  Further details
concerning trajectories and also the phenomenology of the conjugate
variables $\eta$ is provided in App.~\ref{sec:selkov_numerics}.

\section{Concluding remarks}
\label{sec:conclusions}

We have considered the properties of a third representation of the
generator of a stochastic CRN complementing its transition matrix and
Liouville operator, expressed directly in the lattice of factorial
moments of a probability distribution over number states.  This
generator provides a potentially efficient connection between nearby
moments, but suffers the limitation that its boundary conditions are
not simple to handle for high-dimensional systems.

We have shown how a continuum approximation, assuming
slow variation of the ratios of adjacent factorial moments, combines
with information already familiar from momentum-space WKB methods, to
provide a leading exponential estimate for the values of moments
throughout the lattice.  The limit~(\ref{eq:L0_kR_args}), in which the
generator of the moment representation reduces to the Liouville
operator on a stationary $\mathcal{L} = 0$ solution, defines a map
from the orders $k$ of factorial moments to positions along
large-deviation trajectories that control the leading exponential
behavior of moment ratios in neighborhoods of $k$.

Whereas for direct solution of the moment recursion equations,
boundary conditions are difficult to specify where some components of
$k$ become small, in the momentum-space WKB approximation solutions
are propagated to the boundary along rays emanating from fixed points
in the interior, as in the Freidlin-Wentzell approach to computing
boundary values for solutions to diffusions on basins of attraction.
Position-space WKB methods can be used to approximate finite-size
corrections using the simplified scale factors and rate functions
computed from the large-deviations limit.

The moment representation also provides a natural interpretation for
the coherent-state fields $\left( {\xi}^{\dagger} , \xi \right)$ which
appear in Hamilton-Jacobi theory as a canonical transform of the
number-potential fields $\left( \eta , n \right)$ that are the natural
variables for momentum-space WKB methods.  Whereas $n$ is the index,
and $\eta$ the log-derivative, of the stationary distribution, $\xi$
takes a value equal to the ratio $R$ of adjacent factorial moments.
$e^{\eta} = {\xi}^{\dagger}$ is then the ratio of probability flow
rates normal to isoclines of the stationary probability.

The three representations of the generator of a stochastic CRN use the
three classes of random variables as bases: ${\sigma}_{\rn}$ (the
discrete indicator function on states, for the distribution
${\rho}_{\rn}$), ${\rn}^{\underline{k}}$ (discrete powers, for moments
${\Phi}_k$), and $z^{\rn}$ (continuous tilts, for the MGF $\phi \!
\left( z \right)$).  The well-known direct
interconversions~\cite{Doi:SecQuant:76,Doi:RDQFT:76,Peliti:PIBD:85,Peliti:AAZero:86}
and the new one derived
in~\cite{Krishnamurthy:CRN_moments:17,Smith:CRN_moments:17} facilitate
choosing bases with good convergence properties for different classes
of observables of interest.  The connection among the three through
the large-deviations limit derived here, in addition to providing
approximation methods, shows connections of meaning across the
representations that are sometimes not apparent in their more
indirect, detailed interconversions.

\subsection*{Acknowledgments}

DES was supported in part by NASA Astrobiology CAN-7 award NNA17BB05A
through the College of Science, Georgia Institute of Technology, and
by the Japanese Ministry of Education, Culture, Sports, Science, and
Technology (MEXT) through the World Premiere International program.
He also thanks the Physics Department at Stockholm University
for hospitality in May $2016-2019$, when some of
this work was carried out.
\appendix 

\section{Matched asymptotic expansions of the exact recursion for the
one-species autocatalytic CRN} 
\label{sec:autocat_numerics}

For a one-dimensional system, sufficient boundary data is present in
the Liouville operator $\mathcal{L}$ that formally the exact
recursion~(\ref{eq:Glauber_moment_fact_prod}) at steady state is
sufficient to fix a solution.  However, the asymptotic nature of the
recursion makes its application a problem of stabilization in most
cases, with low-order moments of interest being sensitive to
arithmetic precision or truncation errors in the series.  There is
thus a practical use for the WKB
approximation~(\ref{eq:0123_WKB_dist}) in the estimation of
finite-size effects, which we describe here in relation to the exact
recursion and to the large-system limit.

We use as an example an instance of the one-species
CRN~(\ref{eq:cubic_1spec_scheme}) drawn
from~\cite{Krishnamurthy:CRN_moments:17,Smith:CRN_moments:17}, with
parameters ${\bar{k}}_1 = 1$, $k_1 = 14$, $\epsilon = 36$, $k_2 = 49$.
At these parameters the mean-field rate
equation~(\ref{eq:0123_y1A_rate_eqn}) has stable zeros at $n / n_C =
{\bar{x}}_1 = 1$ and ${\bar{x}}_3 = 9$, and a saddle-point zero
${\bar{x}}_2 = 4$.

Traditionally, moment series are solved recursively upward in $k$:
${\Phi}_0 \equiv 1$ by normalization of probability, $R_0 \equiv
{\Phi}_1$ is estimated by the mean-field equations, and so forth.  The
lower-order moments are assumed to be not only the larger components
of the variation, but also those components most robust in a simple
model against parameter imprecision or mis-specification of model
details~\cite{Smith:evo_games:15}.  Therefore, as emphasized
in~\cite{Krishnamurthy:CRN_moments:17,Smith:CRN_moments:17}, it is
counterintuitive that the recursive solution for ${\hat{R}}_k$ is
anchored not at $k \rightarrow 0$ by $x$ in the mean-field solution,
but rather at $k \rightarrow \infty$ by ratios of parameters in
$\mathcal{L}$ that generally differ from the mean of $x$ at any of its
fixed points, and which are related to probability currents that can
arise in networks of nonzero deficiency
(see~\cite{Feinberg:def_01:87,Anderson:product_dist:10} for
elaborations on deficiency).  The downward-going recursion is stable
for $k / \left< \rn \right>$ sufficiently large, so the asymptotic
value can be imposed on $R_k$ at large finite $k$, and truncation
errors are exponentially attenuated as $k$ decreases.  However, the
downward recursion becomes unstable below some characteristic scale
for $k / \left< \rn \right>$, and errors that were initially
attenuated are amplified, along with numerical precision errors, in a
diverging asymptotic expansion that cannot be continued if $n_C$ is
large enough.  

Therefore an upward-going asymptotic expansion must be initiated from
an ansatz at small $k$, and matched to the downgoing expansion at an
intermediate scale where both are marginally stable.  Because the
upward-going solution is stable in increasing $k$, imprecision in the
ansatz is attenuated, but this also means that the matching conditions
to the downgoing asymptotic solution provide weak constraints on the
small-$k$ ansatz at large $n_C$.  It is to substitute for this weak
constraint that the WKB approximation~(\ref{eq:0123_WKB_dist}) can be
useful.

\begin{figure}[ht]
\begin{center} 
  \includegraphics[scale=0.325]{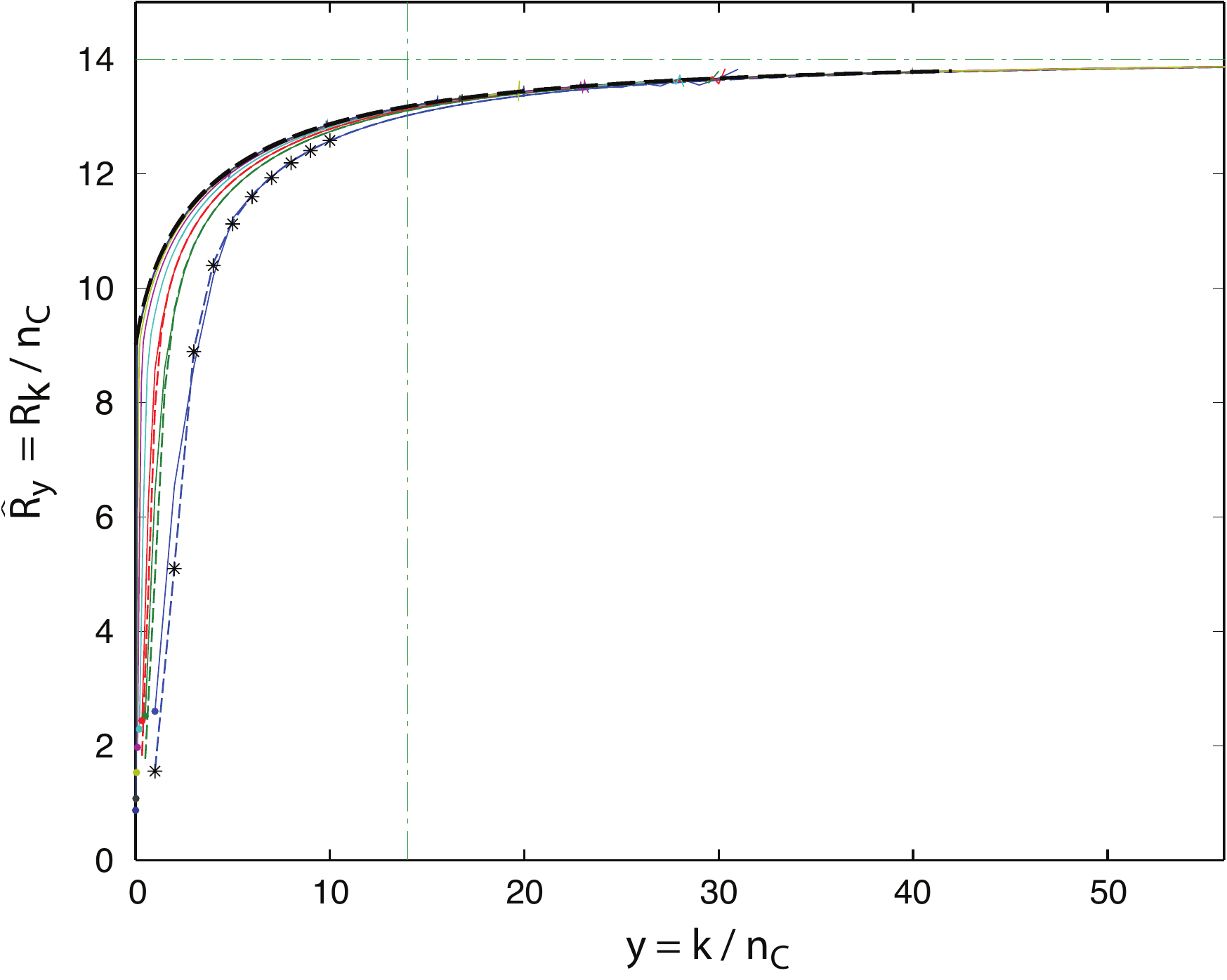} 
 \includegraphics[scale=0.325]{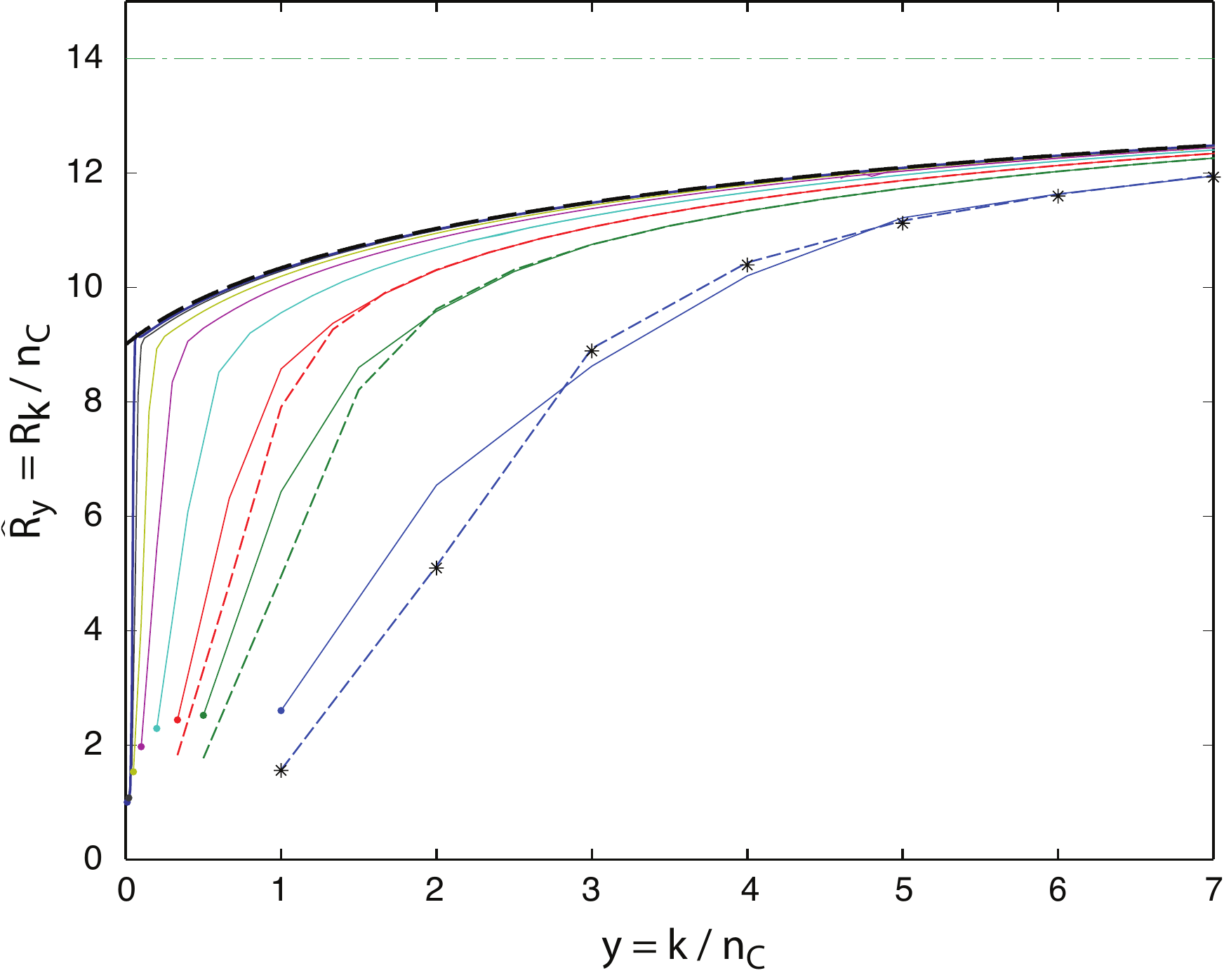} 
  \caption{
  A set of solutions at $n_C = \left\{ 1, 2, 3, 5, 10, 20, 50, 100
  \right\}$ to the exact recursion relations for the one-species
  autocatalytic CRN~(\ref{eq:cubic_1spec_scheme}).  First panel
  illustrates application of the matched asymptotic expansions
  explained
  in~\cite{Krishnamurthy:CRN_moments:17,Smith:CRN_moments:17} in the
  transition region $y \gtrsim 14$ for this model.  Second panel
  expands the abscissa to characterize the WKB
  approximation~(\ref{eq:0123_WKB_fact_moms}).  For $n_C = \left\{ 1,
  2, 3 \right\}$ the upper asymptotic boundary condition extends
  stably to $k = 1$, and agrees closely with the $\hat{R}$ values from
  Gillespie simulations (black asterisks) for $n_C = 1$.  Upward-going
  solutions initiated at values ${\hat{R}}_0$ (colored dots) from the
  WKB approximation are computed at all $n_C$; their deviation from
  the downward solutions gives an estimate of the error from the WKB
  approximation at small $n_C$.  Horizontal dashed green line is the
  upper asymptotic value ${\hat{R}} \le 14$ for these parameters; that
  value is also the characteristic scale for matching of asymptotic
  expansions (vertical green dashed line).  Black-dashed upper curve
  is the higher branch $\hat{R} \! \left( y \right)$ from
  Eq.~(\ref{eq:0123_R_recur_from_CRNLP}).
  \label{fig:scaled_moms_bistable_clean} 
  } 
\end{center}
\end{figure}

Fig.~\ref{fig:scaled_moms_bistable_clean} shows how the WKB starting
conditions for low-order modes relates to full solutions where those
exist, and to Gillespie simulations for one case (at $n_C = 1$).  The
figure shows solutions to the exact recursion
equation~(\ref{eq:rec_0123}) for values $n_C = \left\{ 1, 2, 3, 5, 10,
20, 50, 100 \right\}$.  For the first three of these values, the
recursion is stable all the way to $k = 1$, and at $n_C = 1$ the
low-order modes that can be estimated from simulation are excellent
matches for the recursive solution.  Upward-going solutions, initiated
with $R_0$ and $R_1$ from the WKB approximation are shown for all
eight values of $n_C$, and compared to the downward-going solutions
for $n_C = \left\{ 1, 2, 3 \right\}$.

The figure shows that the difference between the WKB ansatz and the
exact solution is smaller than the separation between adjacent values
of $n_C$, and decreases as $n_C$ grows.  Within each pair of curves
(solid, dashed), attenuation of the error in the WKB initial condition
with increasing $k$ is shown by the rapid convergence of the upgoing
and exact downgoing solutions.  The figure shows also that for $n_C
\sim 100$ the solution is well-approximated by the eikonal solution.
The upward-going recursive solutions only are shown, on logarithmic
scale to expand the transition region between branches of
Eq.~(\ref{eq:0123_R_recur_from_CRNLP}), in
Fig.~\ref{fig:scaled_moms_bistable_good_ks} in the text.

\section{Generator, trajectories, and effective potential for the
two-species cross-catalytic CRN} 
\label{sec:crosscat_numerics}

The cross-catalytic CRN~(\ref{eq:cubic_2spec_scheme}) exactly
recapitulates the large-deviation properties of the autocatalytic
CRN~(\ref{eq:cubic_1spec_scheme}) along the axis of symmetry $x_A =
x_B$, and approximates the solutions to the exact recursions, though
in more complicated form, at finite $n_C$, as derived
in~\cite{Smith:CRN_moments:17}.  Its large-deviation behavior remains
simple in the sense that escape trajectories do not cross (so the
eikonal approximation is valid everywhere), and that they converge
strongly toward time-reverses of solutions to the mean-field equations
of motion~(\ref{eq:CRN_firstmom_EOM}) in neighborhoods of the
diagonal, leading to a computationally efficient approximation of the
effective potential $\hat{\Xi} \! \left( x \right)$ in that region.
Away from the diagonal, however, the irreversible nature of the CRN
becomes visible as escape trajectories deviate from mean-field
trajectories, leading to most-probable excursions that are loops, and
a non-trivial relation between the separatrices for mean-field flow
and for escapes.  Here we provide details of these properties of the
eikonal approximation, retaining the same parameters $\epsilon = 36$,
$k_2 = 49$, $k_1 = 14$, ${\bar{k}}_1 = 1$ used for the 1-species CRN
from App.~\ref{sec:autocat_numerics}.

We begin by noting the exact form of the 2-species generator in the
moment representation.  A set of descaled variables was introduced
in~\cite{Smith:CRN_moments:17}
\begin{align}
  {\omega}_{+} 
& \equiv
  \epsilon / k_1 
& {\omega}_{\Delta} 
& \equiv 
  k_2 / {\bar{k}}_1 - {\omega}_{+} 
\nonumber \\ 
& = 
  \varepsilon n_C^2 / {\kappa}_1
& 
& = 
  k_2 n_C^2 / {\bar{\kappa}}_1 - {\omega}_{+} 
\nonumber \\ 
& \equiv {\hat{\omega}}_{+} n_C^2 
& 
& = 
  {\hat{\omega}}_{\Delta} n_C^2 
\label{eq:omega_eta_def}
\end{align}
The form of the exact generator in the moment hierarchy is 
\begin{widetext}
\begin{align}
\lefteqn{
  - \frac{{\left( \Lambda \Phi \right)}_k}{{\bar{k}}_1}
  = 
  - {\omega}_{\Delta} \left( k_A + k_B \right) {\Phi}_k + 
  {\omega}_{+} 
  \left[
    k_A \left( K_1 {\Phi}_{k-1_a} - {\Phi}_k \right) + 
    k_B \left( K_1 {\Phi}_{k-1_b} - {\Phi}_k \right)
  \right]
} & 
\nonumber \\
& \quad 
  \mbox{} + 
  K_1^2
  \left[ 
    k_A 
    \left( 
      \frac{{\Phi}_{k+1_b}}{K_1} - \frac{{\Phi}_{k+1_a+1_b}}{K_1^2} 
    \right) + 
    k_B 
    \left( 
      \frac{{\Phi}_{k+1_a}}{K_1} - \frac{{\Phi}_{k+1_a+1_b}}{K_1^2} 
    \right) 
  \right] 
\nonumber \\ 
& \quad 
  \mbox{} + 
  K_1
  \left[
    k_A \left( k_A - 1 \right) 
    \left( {\Phi}_{k-1_a+1_b} - \frac{{\Phi}_{k+1_b}}{K_1} \right) + 
    k_A k_B 
    \left( 
      2 {\Phi}_k - \frac{{\Phi}_{k+1_a}}{K_1} -
      \frac{{\Phi}_{k+1_b}}{K_1} 
    \right) + 
    k_B \left( k_B - 1 \right) 
  \left( {\Phi}_{k+1_a-1_b} - \frac{{\Phi}_{k+1_a}}{K_1} \right) 
  \right]
\nonumber \\ 
& \quad 
  \mbox{} + 
  k_A k_B
  \left[ 
    \left( k_A - 1 \right) \left( K_1 {\Phi}_{k-1_a} - {\Phi}_k \right) + 
    \left( k_B - 1 \right) \left( K_1 {\Phi}_{k-1_b} - {\Phi}_k \right) 
  \right] . 
\label{eq:twospec_rull_gen}
\end{align}
Substituting the exact moment ratios $R_{A k}$ and $R_{B k}$ for shifts in
the moment index, Eq.~(\ref{eq:twospec_rull_gen}) becomes
\begin{align}
\lefteqn{
  - \frac{{\left( \Lambda \Phi \right)}_k}{{\bar{k}}_1}
  = 
  \left\{
    - {\omega}_{\Delta} \left( k_A + k_B \right) + 
    {\omega}_{+} 
    \left[
      k_A \left( \frac{K_1}{R_{A k}} - 1 \right) + 
      k_B \left( \frac{K_1}{R_{B k}} - 1 \right)
    \right]
  \right.
} & 
\nonumber \\ 
& \quad 
  \mbox{} + 
  \left. 
    K_1^2
    \left[ 
      k_A \frac{R_{B, k+1_b}}{K_1} 
      \left( 1 - \frac{R_{A, k+1_a+1_b}}{K_1} \right) + 
      k_B \frac{R_{A, k+1_a}}{K_1} 
      \left( 1 - \frac{R_{B, k+1_a+1_b}}{K_1} \right) 
    \right] 
  \right.
\nonumber \\ 
& \quad 
  \mbox{} + 
  \left. 
    K_1 
    \left[
      k_A \left( k_A - 1 \right) 
      \frac{R_{B, k+1_b}}{K_1}
      \left( \frac{K_1}{R_{A, k+1_b}} - 1 \right) + 
      k_A k_B 
      \left( 2 - \frac{R_{A, k+1_a}}{K_1} - \frac{R_{B, k+1_b}}{K_1} \right) + 
      k_B \left( k_B - 1 \right) 
      \frac{R_{A, k+1_a}}{K_1}
      \left( \frac{K_1}{R_{B, k+1_a}} - 1 \right) 
    \right]
  \right.
\nonumber \\ 
& \quad 
  \mbox{} + 
  \left. 
    k_A k_B
    \left[ 
      \left( k_A - 1 \right) \left( \frac{K_1}{R_{A k}} - 1 \right) + 
      \left( k_B - 1 \right) \left( \frac{K_1}{R_{B k}} - 1 \right) 
    \right]
  \right\}
  {\Phi}_k .
\label{eq:twospec_rull_gen_R}
\end{align}
In the large-system substitution of moment
ratios~(\ref{eq:jacquard_walk}), the exact
expression~(\ref{eq:twospec_rull_gen_R}) reduces to
\begin{align}
  - \frac{{\left( \Lambda \Phi \right)}_k}{{\bar{\kappa}}_1 n_C}
& \rightarrow 
  \left\{
    - {\hat{\omega}}_{\Delta} \left( y_A + y_B \right) + 
    \left[ 
      {\hat{\omega}}_{+} + 
      \left( y_A + {\hat{R}}_{A y} \right) 
      \left( y_B + {\hat{R}}_{B y} \right) 
    \right]
    \left[ 
      y_A 
      \left( \frac{{\hat{K}}_1}{{\hat{R}}_{A y}} - 1 \right) + 
      y_B 
      \left( \frac{{\hat{K}}_1}{{\hat{R}}_{B y}} - 1 \right) 
    \right]
  \right\}
  {\Phi}_k , 
\label{eq:twospec_recur_ctm}
\end{align}
\end{widetext}
where ${\kappa}_1/{\bar{\kappa}}_1 \equiv {\hat{K}}_1 = K_1 / n_C$.

The Liouville operator for the CRN~(\ref{eq:cubic_2spec_scheme}) in
number-potential fields is computed by standard
methods~\cite{Smith:CRN_moments:17} to be
\begin{align}
  \hat{\mathcal{L}}
& = 
  \left( 1 - e^{{\eta}_A} \right)
  \left( 
    \epsilon + {\kappa}_1 x_A x_B
  \right) + 
  \left( 1 - e^{- {\eta}_A} \right)
  \left( 
    {\kappa}_2 + {\bar{\kappa}}_1 x_A x_B
  \right) x_A 
\nonumber \\ 
& \mbox{} + 
  \left( 1 - e^{{\eta}_B} \right)
  \left( 
    \varepsilon + {\kappa}_1 x_A x_B
  \right) + 
  \left( 1 - e^{- {\eta}_B} \right)
  \left( 
    {\kappa}_2 + {\bar{\kappa}}_1 x_A x_B
  \right) x_B 
\label{eq:L_two_species_bistable}
\end{align}
Under the mapping~(\ref{eq:main_SP_relns}),
equations~(\ref{eq:twospec_recur_ctm})
and~(\ref{eq:L_two_species_bistable}) are proportional in the same
manner as Eq.~(\ref{eq:onespec_yR_sol}) showed for the 1-species model.

\begin{figure}[ht]
\begin{center} 
  \includegraphics[scale=0.45]{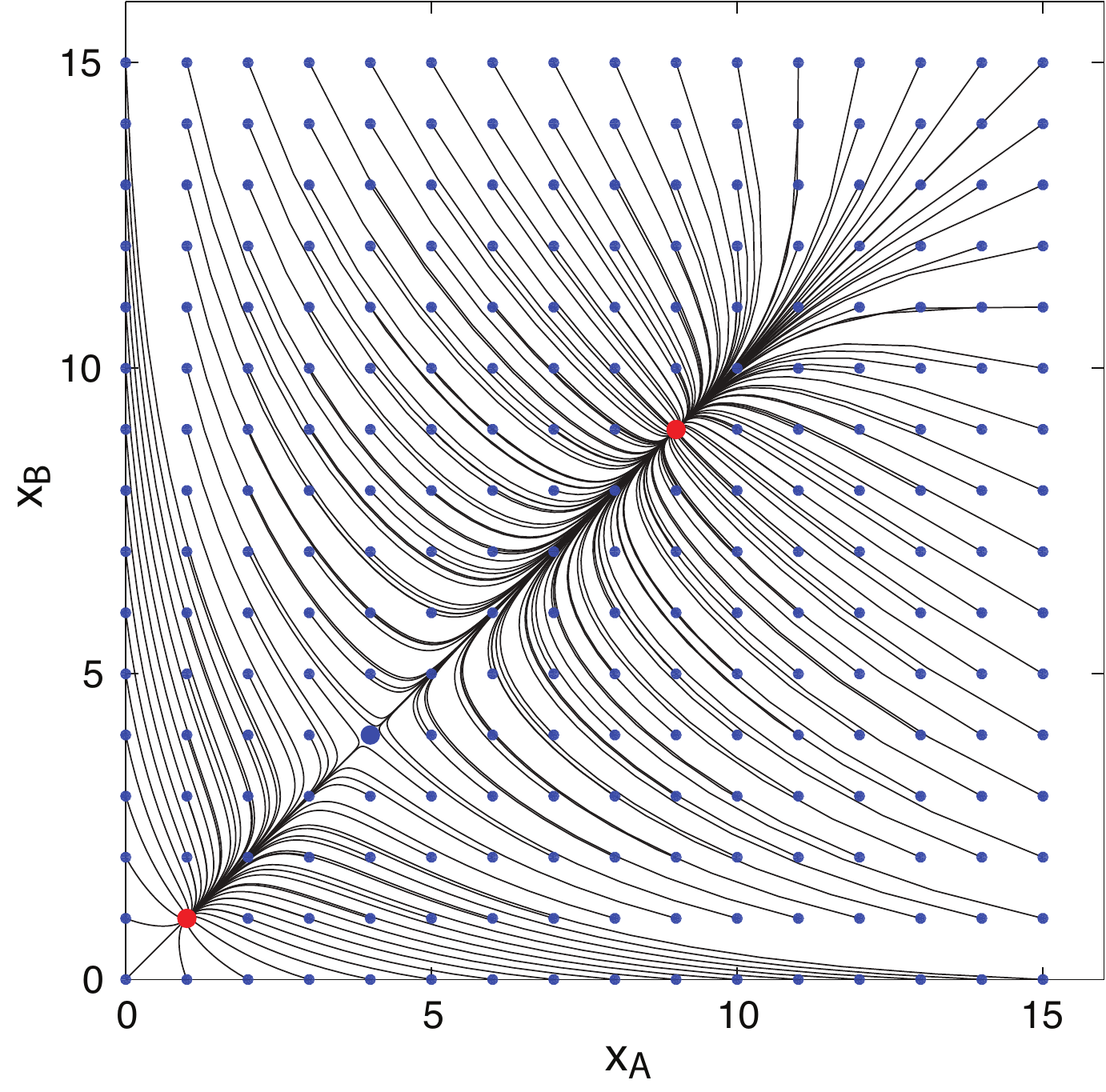} 
  \caption{
  Stationary trajectories at ${\eta}_A = {\eta}_B = 0$ are
  solutions to the mean-field equations~(\ref{eq:rate_ab_cross}).
  Dots indicate starting points of 
  trajectories in this numerical integral.  Curvature toward the axis
  of symmetry $x_A = x_B$ indicates that this axis is a curve of
  slowest departure from the saddle fixed point $\left( 4 , 4
  \right)$, and slowest approach to the attracting fixed points
  $\left( 1 , 1 \right)$ and $\left( 9 , 9 \right)$.
  \label{fig:two_spec_ctm_class_flows} 
  } 
\end{center}
\end{figure}

The general-form eikonal equations that follow from
$\hat{\mathcal{L}}$ in Eq.~(\ref{eq:L_two_species_bistable}) are 
\begin{align}
  d_{\tau} n_A 
& = 
  e^{{\eta}_A}
  \left( 
    \epsilon + k_1 x_A x_B
  \right) - 
  e^{- {\eta}_A}
  \left( 
    k_2 + {\bar{k}}_1 x_A x_B
  \right) x_A 
\nonumber \\ 
  d_{\tau} n_B 
& = 
  e^{{\eta}_B}
  \left( 
    \epsilon + k_1 x_A x_B
  \right) - 
  e^{- {\eta}_B}
  \left( 
    k_2 + {\bar{k}}_1 x_A x_B
  \right) x_B 
\nonumber \\
d_{\tau} {\eta}_A
& = 
  \left( 
    2 - e^{{\eta}_A} - e^{{\eta}_B}
  \right) 
  k_1 x_B + 
  \left( 
    1 - e^{-{\eta}_A}
  \right) 
  \left( 
    k_2 + 2 {\bar{k}}_1 x_A x_B
  \right) 
\nonumber \\ 
& \mbox{} + 
  \left( 
    1 - e^{-{\eta}_B}
  \right) 
  {\bar{k}}_1 x_B^2 
\nonumber \\ 
d_{\tau} {\eta}_B
& = 
  \left( 
    2 - e^{{\eta}_A} - e^{{\eta}_B}
  \right) 
  k_1 x_A + 
  \left( 
    1 - e^{-{\eta}_A}
  \right) 
  {\bar{k}}_1 x_A^2 
\nonumber \\ 
& \mbox{} + 
  \left( 
    1 - e^{-{\eta}_B}
  \right) 
  \left( 
    k_2 + 2 {\bar{k}}_1 x_A x_B
  \right) 
\label{eq:}
\end{align}
Fig.~\ref{fig:two_spec_ctm_class_flows} shows the solutions to these
equations at $\eta \equiv 0$, which are the mean-field
solutions~(\ref{eq:CRN_firstmom_EOM}).  The two basins of attraction
are divided by a separatrix, shown as the blue curve in
Fig.~\ref{fig:two_spec_ctm_escape_sep_bkgnd} in the text.

\begin{figure}[ht]
\begin{center} 
  \includegraphics[scale=0.45]{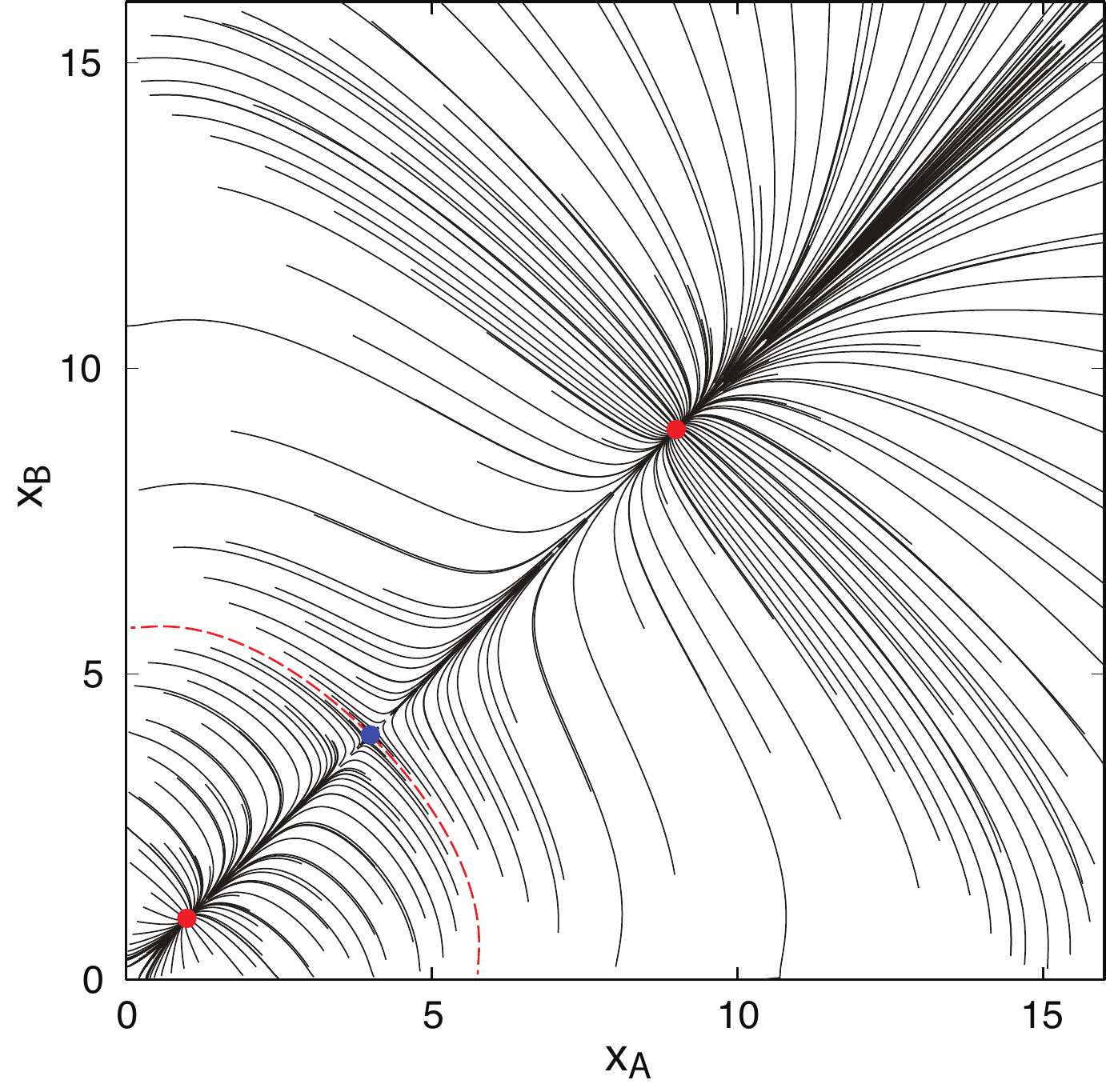} 
  \caption{
  A representative sample of escape rays from the three fixed points.
  For the symmetric model, escape rays converge to the time-reverses
  of solutions to the classical equations of
  motion~(\ref{eq:CRN_firstmom_EOM}) in infinitesimal neighborhoods of
  the fixed points (and along the eikonals from the stable fixed
  points to the saddle.  The escape separatrix falls where rays from
  the lower and upper fixed points become parallel.
  \label{fig:two_spec_ctm_rays_everywhere} 
  } 
\end{center}
\end{figure}

Fig.~\ref{fig:two_spec_ctm_rays_everywhere} shows corresponding escape
trajectories at $\eta \neq 0$, originating in very close neighborhoods
of the stable fixed points.  It can be seen that escapes originating
close to the diagonal, as they approach the saddle, diverge and that
escapes from the two fixed points run parallel.  Their line of
convergence is the escape separatrix shown in dashed red in
Fig.~\ref{fig:two_spec_ctm_rays_everywhere} and also in
Fig.~\ref{fig:two_spec_ctm_escape_sep_bkgnd}.  It is clear that the
escape separatrix is the locus of points with equal probability to be
reached from either fixed point~\cite{Dykman:chem_paths:94}, because
the probabilities on this curve are conditional upon reaching the
saddle, for which the probability is the same from either fixed point
by the construction~(\ref{eq:eff_pot_introd}) of the effective
potential.  Because the conjugate momentum field $\eta$ is continuous
across this curve, $- \int \eta \, dn$ converges to the
log-probability along the separatrix from either side. 

Comparison of Fig.~\ref{fig:two_spec_ctm_class_flows} and
Fig.~\ref{fig:two_spec_ctm_rays_everywhere} shows that escape
trajectories converge to time-reverses of mean-field trajectories in
neighborhoods of the diagonal -- a property that is exact for
1-dimensional systems, noted in the deerivation of
Eq.~(\ref{eq:birth_death_exact}) fn.~\ref{fn:1D_birth_death} -- and
that they are exact reverses along the diagonal, explaining why the
cross-catalytic CRN recovers the large-deviations limit of the
autocatalytic CRN on the axis of symmetry.

However, away from the diagonal escape trajectories are concave toward
the origin, whereas converging trajectories are concave away from the
origin, reflecting the intrinsic irreversibility of the CRN rules.
Two consequences are that excursions from a basin, which relax by
mean-field return, are loops, and also that the escape separatrix is
entirely contained within the basin of attraction of the lower fixed
point ${\bar{x}}_1$.  While off-diagonal excursions are subleading
relative to escapes along the diagonal, and thus do not alter the
balance of probability flows in the large-deviations limit, it follows
that there is a net flow of probability from the fixed point
${\bar{x}}_3$ to the fixed point ${\bar{x}}_1$ by such excursions.

\subsubsection*{Time-reversal approximation to the effective potential}

The eikonal approximation to the full effective potential surface can
be tedious to extract.  The existence of a Liouville theorem for the
dynamical system with ``Hamiltonian'' $\mathcal{L}$ implies that
trajectory integrals have equal numbers of stable and unstable
directions integrated either forward or backward in time, and these
amplify errors in boundary conditions and arithmetical precision
limits exponentially into macroscopic errors.

The assimilation of escapes to time-reverses of converging
trajectories near fixed points, and in the cross-catalytic CRN along
the whole skeleton of eikonals between fixed and saddle points, leads
to an approximation in which $\eta$ is chosen to place $\mathcal{L} =
0$ and to exactly reverse velocity at each point $x$.  Effectively,
one treats probability-current balance along eikonals in higher
dimensions as if it were exact, as it is in one dimension, ignoring
refractive effects on the stationary trajectory.  The resulting $\eta$
field does not satisfy the true eikonal equation for escapes, but for
this CRN the corrections will become small near the diagonal that
determines the main scales of the effective potential.

\begin{figure}[ht]
\begin{center} 
  \includegraphics[scale=0.5]{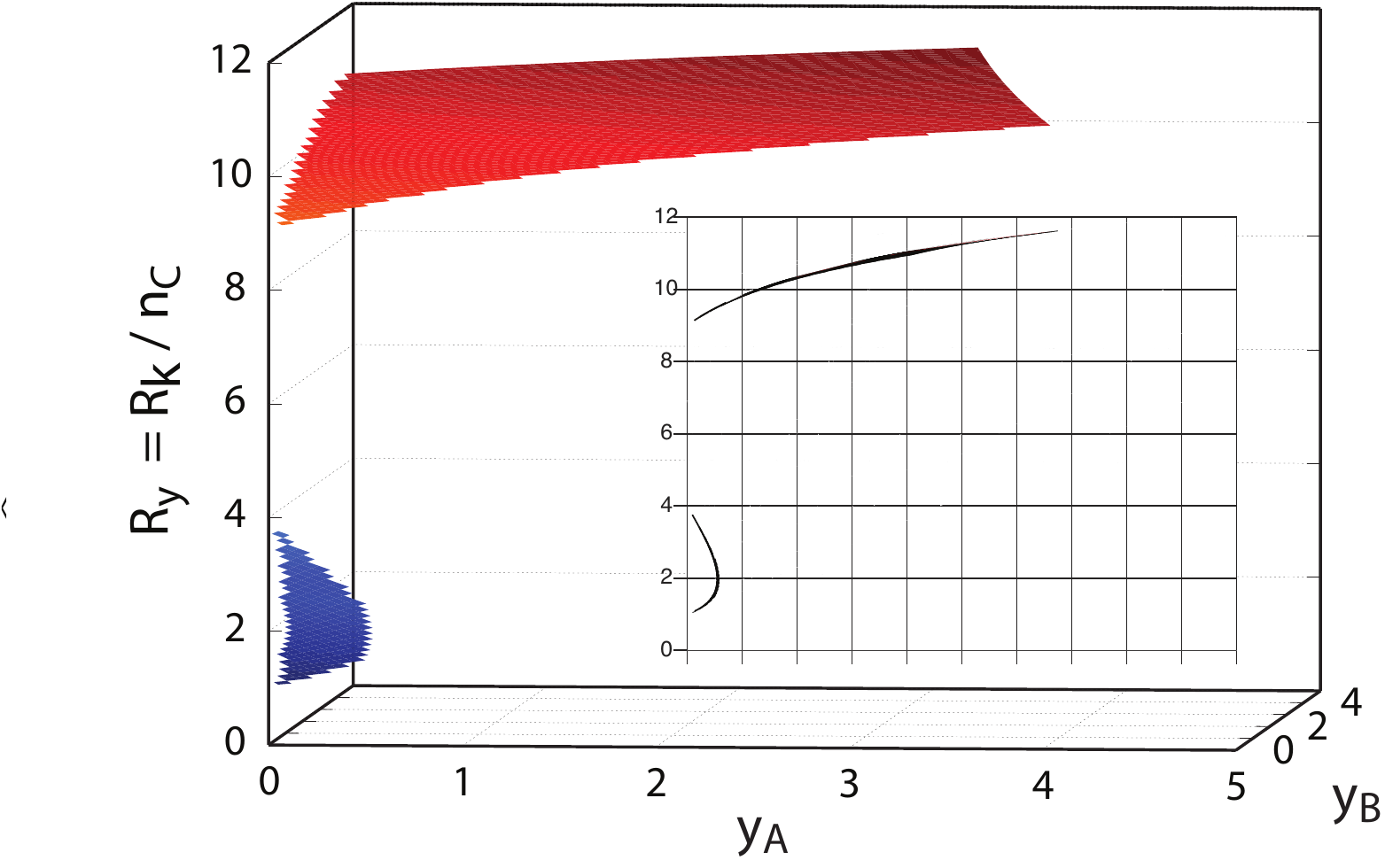} 
  \caption{
  The two branches of the solution $\hat{R} \! \left( y \right)$ in
  the time-reverse approximation~(\ref{eq:twospec_eta_inv_approx}) for
  $\bar{\eta}$.  The branch controlled by the fixed point
  ${\bar{x}}_1$ is shaded blue, and the branch controlled by
  ${\bar{x}}_3$ is shaded red.  Inset shows rotated axes to project
  the surface onto the coordinate $y_A + y_B$, showing that $\hat{R}
  \! \left( y \right)$ is almost independent of $y_A - y_B$.
  \label{fig:two_spec_R_vs_y} 
  } 
\end{center}
\end{figure}

The time-reversing $\eta$ values in the cross-catalytic CRN,
corresponding to the exact solution in the one-variable CRN, are 
\begin{align}
  {\bar{\eta}}_A
& \approx 
  \log 
  \left[
    \frac{
      \left( 
        {\kappa}_2 + {\bar{\kappa}}_1 {\bar{x}}_A {\bar{x}}_B
      \right) {\bar{x}}_A 
    }{
      \left( 
        \varepsilon + {\kappa}_1 {\bar{x}}_A {\bar{x}}_B
      \right) 
    }
  \right] = 
  \log {\bar{x}}_A - \log {\hat{R}}_A
\nonumber \\ 
  {\bar{\eta}}_B
& \approx 
  \log 
  \left[
    \frac{
      \left( 
        {\kappa}_2 + {\bar{\kappa}}_1 {\bar{x}}_A {\bar{x}}_B
      \right) {\bar{x}}_B 
    }{
      \left( 
        \varepsilon + {\kappa}_1 {\bar{x}}_A {\bar{x}}_B
      \right) 
    }
  \right] = 
  \log {\bar{x}}_B - \log {\hat{R}}_B . 
\label{eq:twospec_eta_inv_approx}
\end{align}
The resulting expressions for the moment ratios, corresponding to
Eq.~(\ref{eq:twospec_diag_ray_sol}) for the 1-species model, are 
\begin{align}
  {\hat{R}}_A = 
  {\bar{x}}_A e^{-{\bar{\eta}}_A}
& \approx 
  \frac{
    \left( 
      \varepsilon + {\kappa}_1 {\bar{x}}_A {\bar{x}}_B
    \right) 
  }{
    \left( 
      {\kappa}_2 + {\bar{\kappa}}_1 {\bar{x}}_A {\bar{x}}_B
    \right) 
  }
\nonumber \\ 
  = 
  {\bar{x}}_B e^{-{\bar{\eta}}_B}
& = 
  {\hat{R}}_B .
\label{eq:twospec_R_timerev_approx}
\end{align}
Note that in this approximation ${\hat{R}}_A$ and ${\hat{R}}_B$ are
the same, motivating the equivalence used to solve
Eq.~(\ref{eq:twospec_yb_bdry_sol}).  The stationary-point relation
between $y$, $x$, and $\hat{R}$ from Eq.~(\ref{eq:main_SP_relns})
remains
\begin{equation}
  y_{A, B} = 
  {\bar{x}}_{A, B} - {\hat{R}}_{A, B} \! \left( \bar{x} \right) . 
\label{eq:twospec_poorman_y_Branches}
\end{equation}
Fig.~\ref{fig:two_spec_R_vs_y} shows the two positive branches of the
solution $\hat{R} \! \left( y \right)$ to
conditions~(\ref{eq:twospec_eta_inv_approx},\ref{eq:twospec_poorman_y_Branches}),
coinciding with the solutions~(\ref{eq:0123_two_y_sols}) of the
autocatalytic CRN along the diagonal.  These $\hat{R}$ values are
well-approximated as functions of $y_A + y_B$ over most of the
space. 

The time-reversing $\eta$ values~(\ref{eq:twospec_eta_inv_approx})
give an approximation of the one-argument  effective
potential~(\ref{eq:eff_pot_def}) 
\begin{align}
  \hat{\Xi} \! \left( \bar{x} \right) 
& \approx 
  \int 
  \sum_{p = a, b}
  {\bar{\eta}}_p d{\bar{x}}_p 
\nonumber \\
& = 
  \sum_{p = a, b}
  {\bar{x}}_p 
  \left[ 
    \log {\bar{x}}_p - 1
  \right] - 
  \int 
  \log \hat{R} \! \left( \bar{x} \right) \, 
  d 
  \left( 
    {\bar{x}}_A + 
    {\bar{x}}_B
  \right) . 
\label{eq:twospec_Xi_timerev}
\end{align}
The corresponding result for the two-argument effective
potential~(\ref{eq:0123_xy_eff_Pot}) tilted by $y = \bar{x} - \hat{R}$
becomes  
\begin{align}
  \mathbf{\hat{\Xi}} \! \left( \bar{x} ; y \right) 
& = 
  \hat{\Xi} \! \left( \bar{x} \right) + 
  \sum_{p = a, b}
  \left\{
    {\hat{R}}_p 
    \left[
      \log {\hat{R}}_p  - 1
    \right] - 
    {\bar{x}}_p \left[ \log {\bar{x}}_p - 1 \right] 
  \right\} 
\nonumber \\ 
& = 
  \sum_{p = a, b}
  {\hat{R}}_p 
  \left[ 
    \log {\hat{R}}_p  - 1
  \right] - 
  \int 
  \log \hat{R} \! \left( \bar{x} \right) \, 
  d 
  \left( 
    {\bar{x}}_A + 
    {\bar{x}}_B
  \right) . 
\label{eq:twospec_xy_eff_Pot}
\end{align}
In these two equations, ${\bar{x}}_p \left[ \log {\bar{x}}_p - 1
\right]$ or ${\hat{R}}_p \left[ \log {\hat{R}}_p - 1 \right]$ are
local potential terms that can be computed without computing escape
trajectories.  The other term ($\int \log \hat{R} \! \left( \bar{x}
\right) \, d \left( {\bar{x}}_A + {\bar{x}}_B \right)$) still
depends on a trajectory $\bar{x}$ terminating in the argument of
$\Xi$.  However, because in this approximation ${\hat{R}}_A \approx
{\hat{R}}_B$, only the component $d \left( {\bar{x}}_A + {\bar{x}}_B
\right)$ contributes, and for small $x$, corrections from the
dependence of ${\hat{R}}$ on $\left( {\bar{x}}_A + {\bar{x}}_B
\right)$ remain small. 

\begin{figure}[ht]
\begin{center} 
 \includegraphics[scale=0.5]{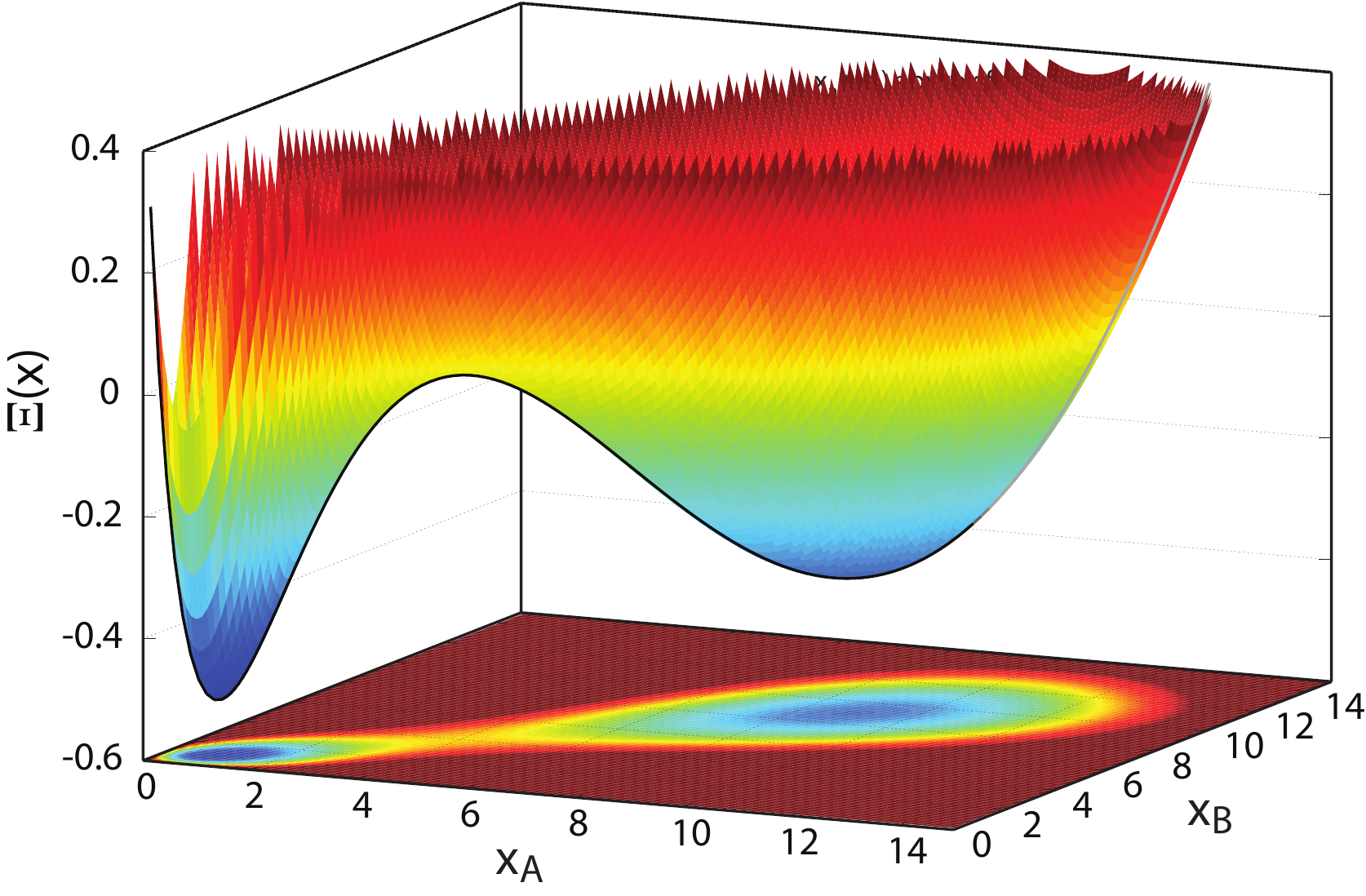} 
  \caption{
  The effective potential $\hat{\Xi} \! \left( \bar{x} \right)$ from
  Eq.~\ref{eq:twospec_Xi_timerev}.  The contour along the diagonal
  ${\bar{x}}_A = {\bar{x}}_B$, equals $2\times\mbox{}$ the value of
  the one-species model from~\cite{Smith:CRN_moments:17}.  Colormap of
  the potential is projected onto the floor of the plot.
  \label{fig:two_spec_pot_rev_flat} 
  } 
\end{center}
\end{figure}

Fig.~\ref{fig:two_spec_pot_rev_flat} shows the approximate
solution~(\ref{eq:twospec_Xi_timerev}).  The black line along the
diagonal equals twice the value from the exact eikonal solution for
the autocatalytic CRN (the cross-catalytic system scales as two
duplicate copies of the autocatalytic system).

\section{Structure of escapes and estimation of the effective
potential for the Selkov model}
\label{sec:selkov_numerics}

The Selkov model was analyzed by Dykman \textit{et
al.}~\cite{Dykman:chem_paths:94} using eikonal methods, and the case
we consider here was chosen to reproduce the major qualitative
features seen in that analysis.  Like the cross-catalytic CRN, the
Selkov CRN has a cubic mean-field equation and one linear steady-state
condition ensuring that all fixed points will lie along a line.  It
generalizes the cross-catalytic CRN, in which the fixed-point line is
by construction the axis of symmetry under $A \leftrightarrow B$, to a
1-parameter degree of freedom setting the slope of the line.  Unlike
the cross-catalytic CRN, where the slope is 1, in the Selkov model the
slope is negative.  Thus the monotonicity of fixed-point values in
$\left( {\bar{x}}_A , {\bar{x}}_B \right)$ is opposite for species $A$
and $B$, leading to a trade-off in the orders $k_A$ and $k_B$ along
the front where control of fluctuations switches from the more-stable
to the less-stable fixed point.

The asymmetric catalytic schema~(\ref{eq:Selkov_scheme}) that gives
the Selkov model a negative slope for its fixed points also leads to
vorticity in the stationary trajectories, which has the same sign
for both classical mean-field equations and escape trajectories.  Thus
excursions that return to a fixed point are even more exaggerated
loops in the Selkov model than in the cross-catalytic CRN.  Escape
trajectories also form caustics, leading to regions in the state space
for which the eikonal approximation becomes invalid, a phenomenon
noted for eikonals in first-passage problems by Maier and
Stein~\cite{Maier:caustics:93} and remarked in this model
in~\cite{Dykman:chem_paths:94}.  

The parameters for the numerical example were chosen as follows: The
free parameter determining the slope in the reaction
schema~(\ref{eq:Selkov_scheme}) we denote as $\left( {\bar{k}}_1 - k_3
\right) / \left( {\bar{k}}_1 + k_3 \right) = \alpha$.  The vorticity
in the mean-field equations~(\ref{eq:CRN_firstmom_EOM}) increases as
$\alpha$ decreases; to study its effect on escapes and caustics we set
$\alpha = -1/3$ in the example.

By choice of the relative normalizations of ${\rn}_A$ and ${\rn}_B$ we
may set $k_2 = {\bar{k}}_2$.  By choice of the absolute normalization
of $\rn$ relative to the buffering species implicit in the Feinberg
null species $\varnothing$, we may set ${\bar{k}}_1 = 1 + \alpha =
2/3$, $k_3 = 1 - \alpha = 4/3$. 

To preserve as much similarity as possible to the previous two
examples in which the fixed point numbers ${\bar{x}}_1 : {\bar{x}}_2 :
{\bar{x}}_3$ were in the ratio $1:4:9$, we seek fixed points with the
ratios $\left( {\bar{x}}_3 - {\bar{x}}_2 \right) : \left( {\bar{x}}_2
- {\bar{x}}_1 \right) = 5:3$.  The smallest integer-valued fixed
points with this property are given by assignments: $k_2 = \left( 1 -
{\alpha}^2 \right) / \left( 2 \times {14}^2 \right) \approx 0.0023$, 
${\bar{k}}_3 = 2 \left( 1 - \alpha \right) \times \left( 36/49 \right)
\approx 1.9592$, $k_1 = 4 \times 14 - {\bar{k}}_3 \approx 54.041$.
The resulting three solutions for the fixed points are
$ \left( {\bar{x}}_A , {\bar{x}}_B \right) \in 
  \left\{ 
    \left(
      80 , 2 
    \right) , 
    \left(
      68 , 8 
    \right) , 
    \left(
      48 , 18
    \right) 
  \right\} $.  

\begin{figure}[ht]
\begin{center} 
  \includegraphics[scale=0.45]{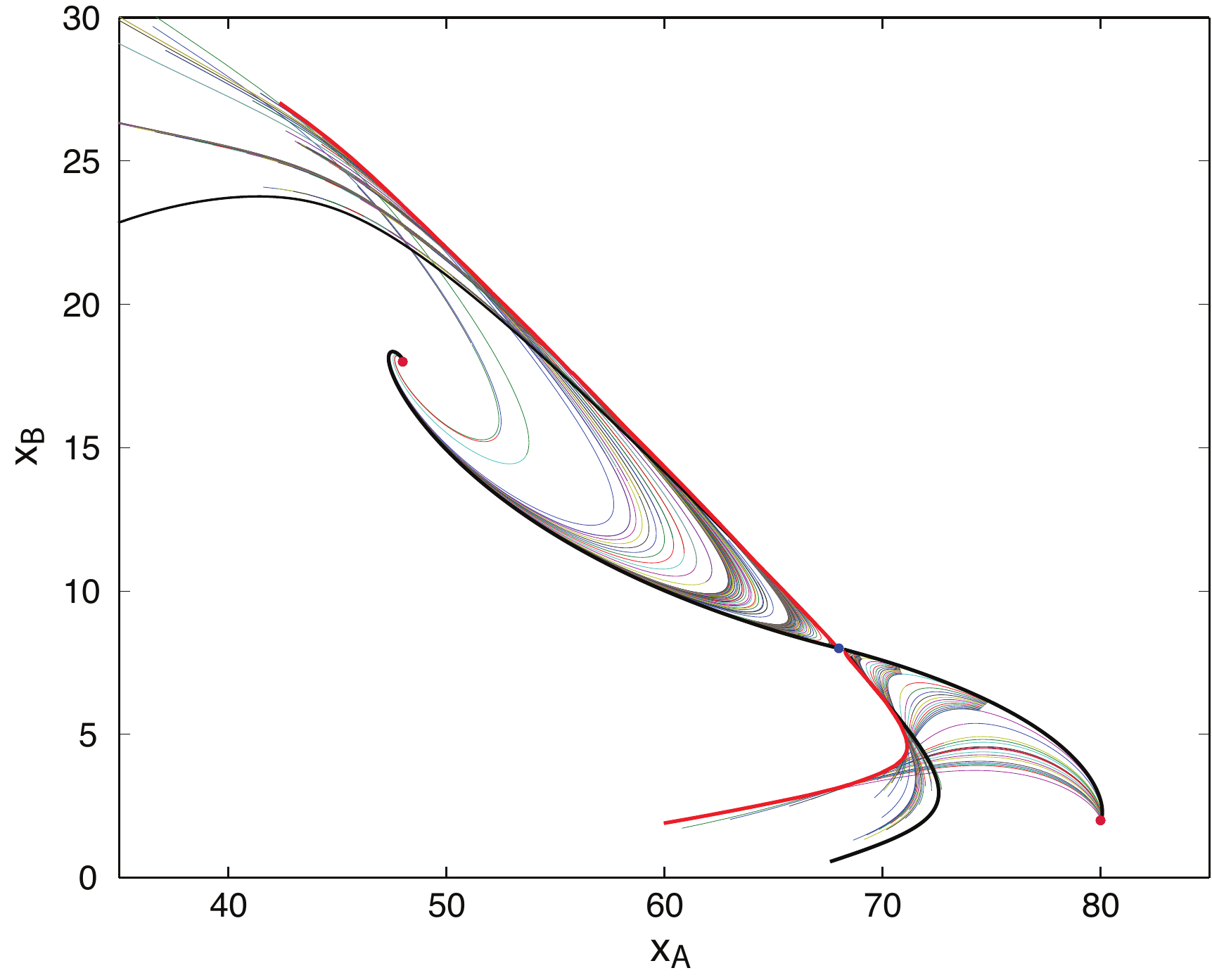} 
  \caption{
  Sets of escape rays from either stable fixed point (thin stroke),
  bounded on one side by the eikonal from the fixed point to the
  saddle (heavy black), and on the other by a caustic (heavy red).
  Rays departing from the eikonal very near the saddle converge toward
  limiting rays (heavy black).
  \label{fig:selkov_rays_caustics} 
  } 
\end{center}
\end{figure}

Solutions to the mean-field equations at these parameters are shown in
Fig.~\ref{fig:selkov_ctm_class_flows}, and those can be compared
to solutions for escape trajectories in
Fig.~\ref{fig:selkov_rays_caustics}.  In the second figure, we show
only rays exiting counterclockwise to the eikonals from the fixed
points that reach the saddle, to highlight the formation of caustics
by rays that double back before reaching the saddle.  The skeleton of
boundary rays from this figure is plotted alone in
Fig.~\ref{fig:selkov_eiks_limrays} in the text.

Integration of $\int \eta \, dn$ along the eikonals from the stable
fixed points to the saddle gives the cut through the effective
potential $\hat{\Xi} \! \left( x \right)$ along its minimizing contour
shown in Fig.~\ref{fig:selkov_SaddleEiks_eff_pot}.  The profile is
qualitatively similar to that shown in
Fig.~\ref{fig:two_spec_pot_rev_flat} for the cross-catalytic CRN that
these Selkov-model parameters were chosen to resemble.  From the
relative minima of $\hat{\Xi} \! \left( x \right)$ we may evaluate the
WKB approximation~(\ref{eq:0123_WKB_fact_moms}) for the domains where
the deeper fixed point ${\bar{x}}_1 = \left( 80 , 2 \right)$ and the
shallower fixed point ${\bar{x}}_3 = \left( 48 , 18 \right)$ control
moment ratios.  The separatrix between these two is estimated to lie
along the linear front
\begin{align}
  \begin{array}{c}
    \left[ 
      \begin{array}{cc}
        k_A & k_B
      \end{array}
    \right] \\
    \phantom{\mbox{}}
  \end{array}
  \left[ 
    \begin{array}{c}
      \log \left( 48 / 80 \right) \\
      \log \left( 18 / 2 \right) 
    \end{array}
  \right] 
& = 
  n_C 
  \left( 
    \hat{\Xi} \! \left( {\bar{x}}_3 \right) - 
    \hat{\Xi} \! \left( {\bar{x}}_1 \right) 
  \right) 
\nonumber \\ 
  \approx 
  \begin{array}{c}
    \left[ 
      \begin{array}{cc}
        k_A & k_B
      \end{array}
    \right] \\
    \phantom{\mbox{}}
  \end{array}
  \left[ 
    \begin{array}{c}
      \log \left( 3 / 5 \right) \\
      2 \log \left( 3 \right) 
    \end{array}
  \right] 
& \approx 
  n_C
  \left( 
    0.4462 - 0.1606
  \right) .
\label{eq:selkov_twomin_cross}
\end{align}

\begin{figure}[ht]
\begin{center} 
  \includegraphics[scale=0.45]{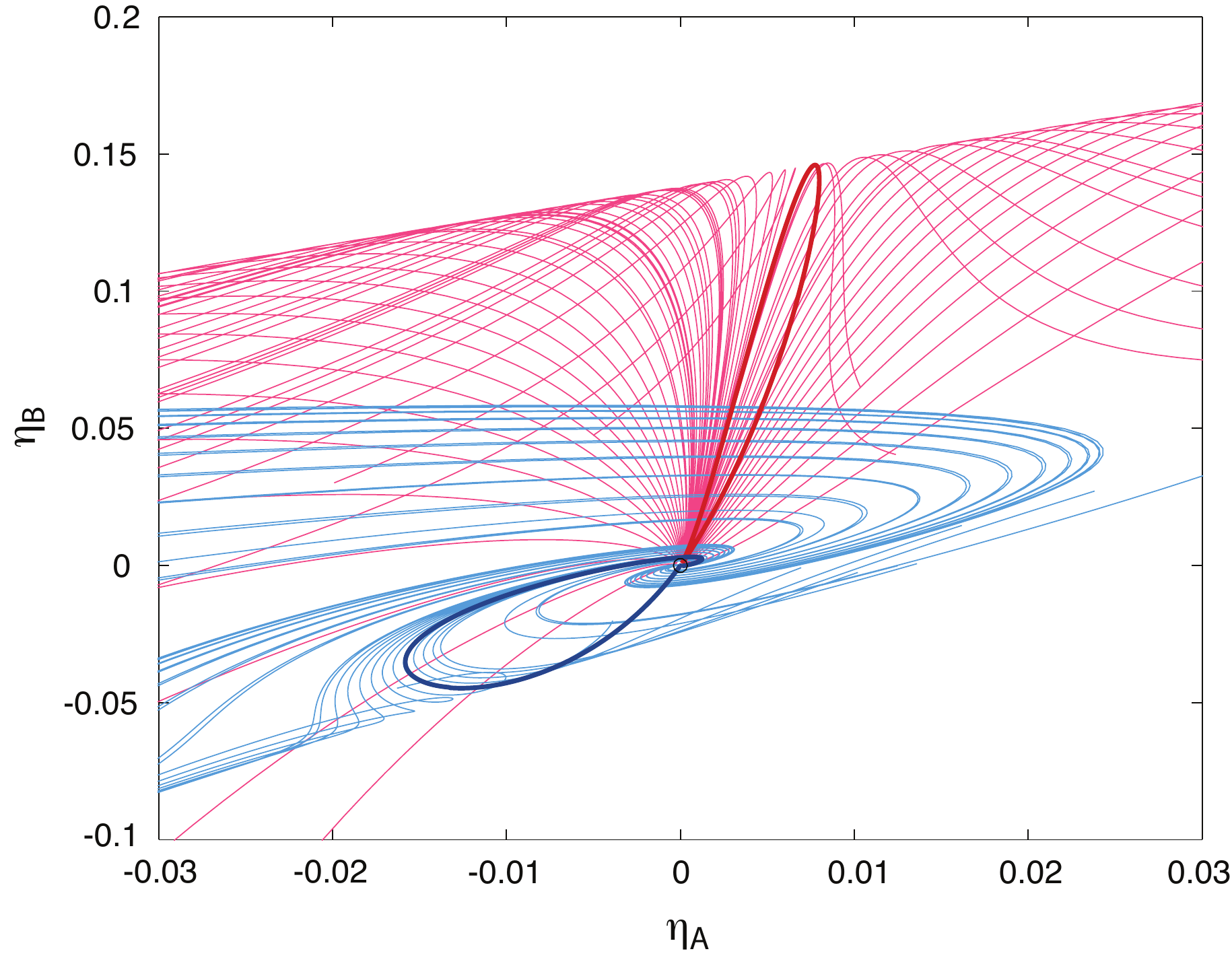} 
  \caption{
  Field values $\eta$ along escape contours from the deeper
  fixed point $\left( 80 , 2 \right)$ in the potential of
  Fig.~\ref{fig:selkov_SaddleEiks_eff_pot} (red) and the shallower
  fixed point $\left( 48 , 18 \right)$ (blue).  $\eta$ values along
  eikonals from the stable fixed points to the saddle are heavy lines.
  $\eta$ integrated along escape contours reaches a maximum in the
  component $\sim \left[ \; -1 \; 1 \; \right]$, along all rays in a
  neighborhood of the eikonal.  Hence $\int dn  \eta$ giving the
  effective potential attains its maximum roughly along this horizon,
  in contrast to the maximum for the cross-catalytic CRN shown in
  Fig.~\ref{fig:two_spec_R_vs_y} (inset), which occurs on a front of
  roughly constant $y_A + y_B$. 
  \label{fig:selkov_eta_horizons} 
  } 
\end{center}
\end{figure}

Fig.~\ref{fig:selkov_eta_horizons} shows the behavior of the potential
field $\eta$ along the two eikonals from the fixed points to the
saddle and along neighboring escape trajectories.  Consistent with the
linear front~(\ref{eq:selkov_twomin_cross}) across which the dominant
contribution to moments is expected to switch from the deeper to the
shallower fixed point, $\eta$ is bounded by roughly linear horizons
along rays escaping from the lower and upper fixed points,
corresponding respectively to the maxima and minima of the cubic
solution~(\ref{eq:twospec_diag_ray_sol}) in the one-dimensional model.
The upper horizon for $\eta$ from the fixed point $\left( 80 , 2
\right)$, with normal $\approx \left[ \; -1 \; 1 \; \right]$
corresponds to the maximum of $\eta$ along the lower branch solution
shown in the inset to Fig.~\ref{fig:two_spec_R_vs_y} for the
cross-catalytic model.  

\begin{figure}[ht]
\begin{center} 
  \includegraphics[scale=0.4]{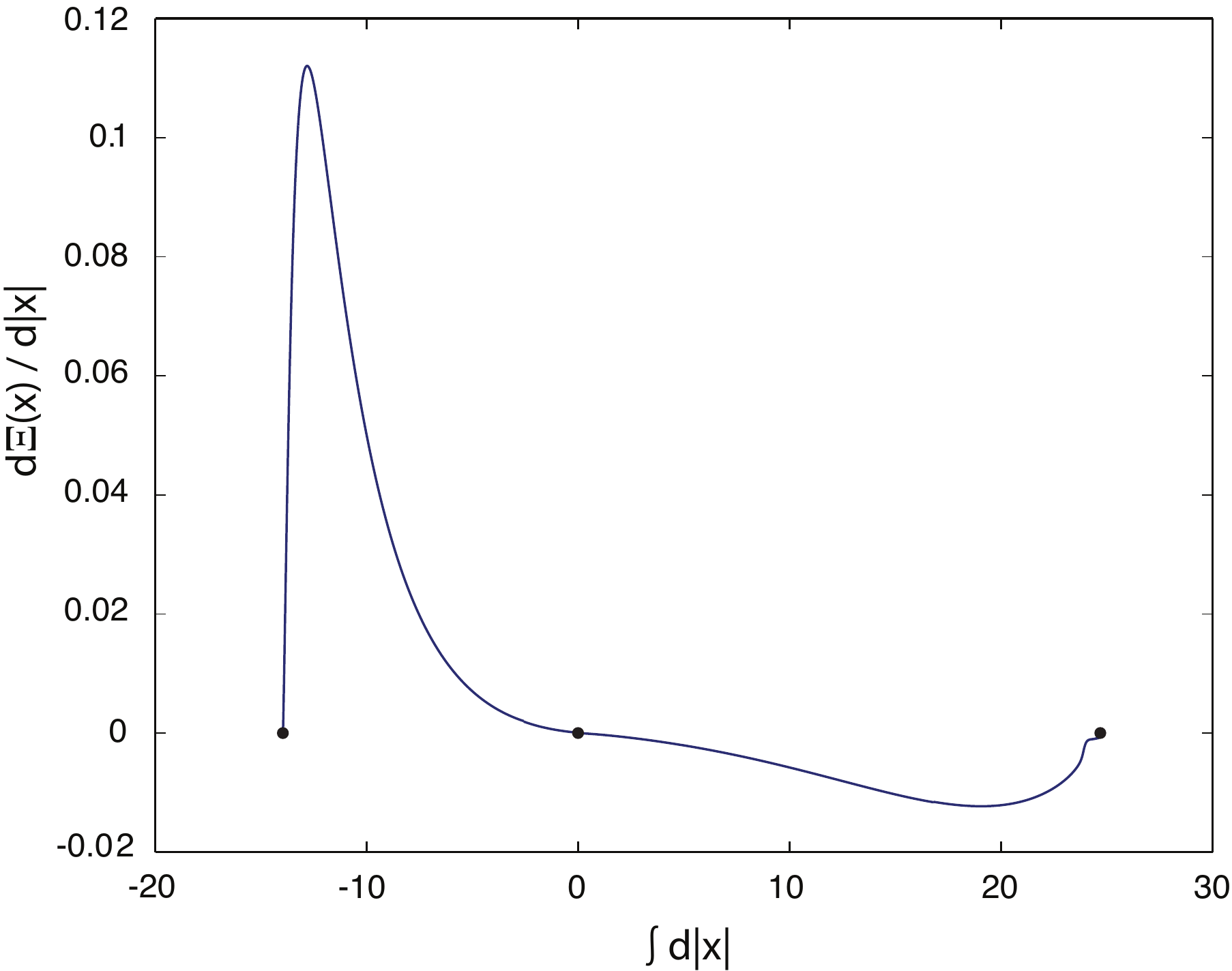} 
  \caption{
  $d \Xi / d \left| n \right|$ versus distance $\int d \left| n
  \right|$ along the eikonals from the two fixed points to the saddle,
  giving the component of $\eta$ along the least improbable escape
  path.  $\eta$ approaches zero linearly near the lower fixed point
  $\left( 80 , 2 \right)$, but its behavior is more complex near the
  upper fixed point $\left( 48 , 18 \right)$, due to the spiral exit
  of the escape ray shown in Fig.~\ref{fig:selkov_eiks_limrays}.  The
  maximum value along the escape from the fixed point $\left( 80 , 2
  \right)$ is similar to that for the cross-catalytic model, and
  comparable to the maximum on the horizon in
  Fig.~\ref{fig:selkov_eta_horizons}.
  \label{fig:selkov_SaddleEiks_eff_pot_grad} 
  } 
\end{center}
\end{figure}

Fig.~\ref{fig:selkov_eta_horizons} shows also that $\eta \rightarrow
0$ on the approach to either stable or saddle fixed points, making a
smooth continuation with the $\mathcal{L} \equiv 0$ condition
satisfied along all mean-field trajectories.  Thus, even though the
curvature of $\hat{\Xi} \! \left( x \right)$ in the neighborhood of
the deeper fixed point is not readily visible in
Fig.~\ref{fig:selkov_SaddleEiks_eff_pot} as it is for the
cross-catalytic model in Fig.~\ref{fig:two_spec_pot_rev_flat}, the
potential is quadratic in neighborhoods of all of its fixed points.
Fig.~\ref{fig:selkov_SaddleEiks_eff_pot_grad} shows the gradient $d
\Xi / d \left| n \right|$ (defined as $\eta \, dn / dn^2$
for the line element along the contour) equivalent to
Eq.~(\ref{eq:0123_xy_spinodal}) along the tangent to the eikonals
connecting the stable fixed points to the saddle.  The gradient
remains monotonic in neighborhoods of either fixed point, but its
derivative is not monotone near the fixed point $\left( 48 , 18
\right)$, as a result of the vorticity and shear of all stationary
trajectories in this region.

\vfill
\eject


\end{document}